\documentclass[11pt]{article}
\usepackage{graphicx,epsfig}
\def\Sec{Section}

\def\lesssim{\mathrel{\hbox{\rlap{\hbox{\lower4pt\hbox{$\sim$}}}\hbox{$<$}}}}
\def\gtrsim{\mathrel{\hbox{\rlap{\hbox{\lower4pt\hbox{$\sim$}}}\hbox{$>$}}}}
\def\asim{\mathrel{\hbox{\rlap{\hbox{\lower4pt\hbox{$-$}}}\hbox{$\sim$}}}}
\def\gmcmc{gm$\;\mbox{cm}^{-3}$~}
\def\cmsgm{cm$^2\mbox{gm}^{-1}$~}
\def\cms{cm$\;\mbox{sec}^{-1}$~}
\def\dncms{dyne$\;\mbox{cm}^{-2}$~}
\def\Kelv{$^\mathrm{o}\mbox{K}$~}
\def\ergsdeggm{ergs$\;\mbox{deg}^{-1}\;\mbox{gm}^{-1}$}
\begin{document}

\title{The Properties of Matter\\ in White Dwarfs and Neutron Stars}
\author{Shmuel Balberg and Stuart L.~Shapiro\footnote{also Department of 
Astronomy and National Center for Supercomputing Applicaions, 
University of Illinois at Urbana-Champaign, 
Urbana, IL 61801}\\
\small Department of Physics, University of Illinois at Urbana-Champaign,\\
\small 1110 W.~Green St., Urbana, IL 61801}
\date{}

\maketitle

\begin{abstract}
White dwarfs and neutron stars are stellar objects with masses comparable 
to that of our sun. However, as the endpoint stages of stellar 
evolution, these objects do not sustain any thermonuclear burning
and therefore can no longer support the gravitational load of their own 
mass by generating thermal pressure. Rather, matter in their interiors is 
compressed to much higher densities than commonly found in normal stars, and 
pressure is created by degenerate fermion kinetic energy and particle 
interactions.
As a result, white dwarfs and neutron stars offer unique cosmic 
laboratories for studying matter at very high densities. In this review we 
discuss the basic properties of condensed matter at extreme densities and 
summarize the extent to which these 
properties can be examined by observations of compact objects.
\end{abstract}


\section{Introduction}

Astronomical phenomena provide many examples where matter exists in 
extreme conditions not found in terrestrial environments. 
One example is the high density of degenerate matter 
in ``compact objects'' - the relics of stars that have 
ceased burning thermonuclear fuel, and thereby no longer generate thermal 
pressure to support 
themselves against gravitational collapse. 
By contracting appreciably from their original sizes, the interiors of 
compact objects reach sufficiently high densities to produce 
nonthermal pressure via degenerate 
fermion pressure and particle interactions. Compact objects provide 
cosmic laboratories for studying the properties of matter at high densities. 

Firm observational evidence and 
well-founded theoretical understanding both exist for two 
classes of compact objects which support themselves against 
collapse by cold, degenerate fermion pressure: 
{\bf white dwarfs}, whose interiors resemble a very dense solid, 
with an ion lattice surrounded by degenerate electrons, and 
{\bf neutron stars}, whose cores resemble a giant atomic nucleus -  
a mixture of interacting nucleons and electrons, and possibly other 
elementary particles and condensates. White dwarfs are supported by the 
pressure of degenerate electrons, while neutron stars are supported by 
pressure due to a combination of nucleon degeneracy and nuclear interactions. 
These unique states of matter are achieved by significant 
compression of stellar material. Table~\ref{tab:sizes} compares the principal 
physical quantities of a typical white dwarf and neutron star with those 
of the sun\footnote{Throughout this review we will be using units which 
are the standard in astrophysical research: cgs for microscopic properties 
of matter and solar units (denoted by $\odot$) for macroscopic properties 
of astronomical objects.}. 

\begin{table}[h] 
\caption{Parameters for the sun and a typical white dwarf and neutron star. 
\label{tab:sizes}}
\begin{center}
\begin{tabular}{c c c c c c}
object & mass                & radius & mean density & mean pressure & 
$GM/Rc^2\;\;^{b}$ \\
    & ($M_\odot\;\;^{a}$) &  (km)  & (\gmcmc)  & (\dncms) &     \\ \hline
Sun          
    & 1          & $\sim 7\times 10^5$ & $\sim 1$     & $\sim 10^{12}$  
& $\sim 10^{-6}$    \\
White Dwarf 
    & $\leq 1.4$ & $\sim 5\times 10^3$ & $\sim10^7$   & $\sim 10^{24}$ 
& $\sim 10^{-4}$\\
Neutron Star 
    &   1-3   & $\sim 10$           & $\sim10^{14}$ & $\sim 10^{34}$ 
& $\sim 10^{-1}$ \\
\end{tabular}
\end{center}
$^{a}$ - $M_\odot\equiv 1.989\times10^{33}\;$gm = one solar mass.\\
$^{b}$ - This ratio measures the importance of relativistic gravitation, 
i.e., general relativity.  

\end{table}

Condensed matter in compact objects spans an enormous range of densities, 
which we loosely refer to as ``high densities''.  
These extend from about $7\:$\gmcmc (e.g., the density of terrestrial 
$^{56}_{26}$Fe), 
at the surface of a cold neutron star or white dwarf, 
to as much as $\rho\approx 10^{15}\;$\gmcmc, several times the density 
in atomic nuclei, in the cores of neutron stars. 
Matter at the various densities found in compact objects exhibits a variety 
of novel properties. Electromagnetic, strong, and 
weak interactions all play an important role in determining the character  
of compact objects. Since these objects are bound by gravity, they are 
a meeting point of all four of the fundamental forces of nature. 
Correspondingly, the astrophysics of white dwarfs and neutron stars 
incorporates a wide variety of physics including nuclear, particle, 
solid state and gravitation physics, to name a few areas. 

In this review we briefly survey the theory of condensed matter at high 
densities in compact objects and illustrate how the basic theory is tested 
through astronomical observations. Since we must cover fifteen 
orders of magnitude in density, 
our presentation is at most introductory in nature, and we encourage 
the interested reader to pursue the cited references. A detailed 
introduction to the physics of high density matter and compact objects can be 
found in the textbook {\it Black Holes, White Dwarfs and Neutron Stars: 
the Physics of Compact Objects}, by Shapiro and Teukolsky \cite{ShaTeu83}.

We begin by considering the fundamental nature of cold ($T=0$) 
high density matter in  \Sec~\ref{sect:EoS}, and distinguish between 
different regimes of high density.  
In \Sec~\ref{sect:WDNSconf} we connect these microscopic properties 
with the fundamental macroscopic parameters of a compact object through 
the hydrostatic equilibrium dependence of mass and radius on central 
density. We summarize the fundamental predictions regarding the 
structure of white dwarfs and neutron stars. In 
\Sec~\ref{sect:WDs}-\ref{sect:NSs} we examine how observations of 
white dwarfs and neutron stars can be used to probe the 
properties high density matter. We briefly discuss the perturbative effects 
of a finite temperature in \Sec~\ref{sect:Therm}. 
          
\section{The Cold, High Density Equation of State}
\label{sect:EoS}

The pressure and energies in white dwarfs and neutron stars are nonthermal; 
thermal effects due to a finite-temperature can be treated as a 
perturbation. We may therefore treat high density matter as having a zero 
temperature to very good approximation. The equation of state (EoS) 
of the matter then reduces to a single-parameter function, 
$P(\rho_0)$ and $\rho(\rho_0)$, or $P(\rho)$, where $P$ is the pressure, 
$\rho_0$ is the rest-mass density and $\rho$ is the total mass-energy 
density which accounts for internal (possibly relativistic) 
particle energies as well as rest-mass energy. 

There exist two main regimes of high density, distinguished as follows. 
As long as all nucleons are confined to nuclei, their contribution 
to the total pressure is negligible compared to that of the degenerate 
electrons. At some threshold density, $\rho_{n-drip}$, it becomes favorable 
for the nuclei to disintegrate, i.e., neutrons ``drip'' out of the nuclei 
and form a ``nucleon gas''. 
The standard EoS of Baym, Pethick and Sutherland \cite{BPS}
suggests that $\rho_{n-drip}\approx 4\times 10^{11}\;$\gmcmc. We may 
distinguish between the EoS for 
$\rho\leq\rho_{n-drip}$ which characterizes matter in white dwarfs and in 
the outermost layers of neutron stars, and $\rho>\rho_{n-drip}$, which 
describes matter in the interior of neutron stars. 

\subsection{The Equation of State below Neutron Drip Density: 
$\rho\lesssim 4\times 10^{11}\;$\gmcmc}   

In matter below the neutron drip density the 
ions provide a Coulomb lattice of point-like charges, which 
is (to good approximation) independent of the properties of the surrounding 
electrons. The EoS of such matter is governed mainly by the electron 
gas. To lowest approximation, we may treat the electrons as 
an ideal fermion gas, incorporating some Coulomb corrections 
at relatively low density ($\rho\leq10^4\;$\gmcmc) and corrections due to 
inverse $\beta-$decay just below the neutron drip density
($10^9\;$\gmcmc$\leq\rho\leq\rho_{n-drip}$).
The EoS of condensed matter below neutron 
drip density is well understood. The standard 
equations for cold, degenerate matter in white dwarfs (helium, carbon, 
oxygen and, possibly iron dominated models) have been derived by Chandrasekar 
\cite{ChandraWD} and Salpeter 
\cite{Sal61}, whereas 
models for equilibrated matter\footnote{The equilibrium isotope of matter 
is the nucleus of highest binding energy per nucleon. At low densities this 
isotope is normal $^{56}_{26}$Fe, but as density increases so do the atomic 
mass and neutron to proton ratio \cite{ShaTeu83}.},
are based on the works of Dirac \cite{Dirac} 
and Feynman, Metropolis and Teller \cite{FMT} for 
$\rho\leq 10^4\;$\gmcmc, and e.g., of Harrison and Wheeler \cite{HW} and 
Baym, Pethick and Sutherland \cite{BPS} for higher densities.  

\subsubsection{The Ideal Fermion Gas}

For almost the entire range of high densities the electrostatic energy 
associated with the structure of matter is much smaller than the 
Fermi energies. Consequently, Coulomb forces are generally negligible in a 
first order treatment of the high density EoS. The electron component of 
high density matter can therefore be described by a 
cold, single species gas of noninteracting fermions. 
At zero temperature the fermions fill all the 
states with momentum $p\leq p_F$ and none of the states with $p>p_F$, where 
$p_F$ is the Fermi momentum. The corresponding Fermi energy of the 
particle species is
\begin{equation}\label{eq:E_F}
E_F\equiv \left((p_F c)^2 + (m c^2)^2\right)^{1/2}\;,
\end{equation}
where $c=3\times10^{10}\;$\cms is the speed of light in vacuum and $m$ is the 
fermion rest mass.

For electrons, their number density, $n_e$, is directly related to their 
Fermi momentum, $p_{F,e}$ by integrating over all occupied 
phase space ($h=6.63\times 10^{-27}\;$ergs~sec is Plank's constant): 
\begin{equation}\label{eq:n_fer}
n_e=\int_0^{p_{F,e}}n_e(p)d\!p\equiv
2\frac{1}{h^3}\int_0^{p_{F,e}}{4\pi p^2 d\!p} = 
\frac{8 \pi p_{F,e}^3}{3 h^3}\;.
\end{equation}
The factor of 2 in the second equation arises from the 
electron spin degeneracy. 
 
The pressure the electrons supply is calculated through the mean momentum 
flux of the electron gas,   
\begin{equation}\label{eq:P_fer}
P=\frac{1}{3}\int_0^{p_{F,e}}v_e(p)n_e(p)p d\!p = 
\frac{2}{h^3}\int_0^{p_{F,e}}
\frac{p^2c^2}{(p^2c^2+(mc^2)^2)^{1/2}}4\pi p^2 d\!p = 
\frac{m_e c^2}{\lambda_e^3}\phi(x)\;,
\end{equation}
where $x_e\equiv p_{F,e}/m_e c$ is the electron ``relativity parameter'',  
$\lambda_e\equiv h/(2\pi m_e c)$ is the electron Compton wavelength, 
$v_e=p_e c^2/E_e$ is the electron velocity and
\begin{equation}\label{eq:psiofx}
\phi(x)=\frac{1}{8\pi^2}\left\{x(1+x^2)^{1/2}(2x^2/3-1)+
\ln\left[x+(1+x^2)^{1/2}\right]\right\}\;.
\end{equation}

The mass-energy density of the free electrons is also uniquely related to 
the Fermi momentum as
\begin{equation}\label{eq:e_fer}
\varepsilon_e=\int_0^{p_{F,e}}E_e(p) n_e(p) d\!p = 
\frac{2}{h^3}\int_0^{p_{F,e}}
(p^2c^2+(mc^2)^2)^{1/2}4\pi p^2 d\!p = 
\frac{mc^2}{\lambda_e^3}\chi(x_e)\;;
\end{equation}
where 
\begin{equation}\label{eq:chiofx}
\chi(x)=\frac{1}{8\pi^2}\left\{x(1+x^2)^{1/2}(1+2x^2)-
\ln\left[x+(1+x^2)^{1/2}\right]\right\}\;.
\end{equation}
However, even when the degenerate electrons contribute most of the pressure, 
the mass-energy density is dominated by the rest mass of the ions, which are 
very nonrelativistic at these densities. Thus, the density of the matter 
may be simply taken as 
\begin{equation}\label{eq:rhoWD}
\rho=\rho_0=\frac{n_e m_B}{Y_e}\;,
\end{equation}
where $Y_e$ is the mean number of electrons per nucleon 
and $m_B$ is the mean nucleon mass. In the case of white dwarfs 
we may set $Y_e=Z/A=0.5$ ($Z=$atomic number, $A=$atomic weight) 
which is appropriate for fully ionized helium, carbon or oxygen, 
the most abundant constituents in a white dwarf, and 
$m_B=1.66\times 10^{-24}\;$gm. Combining Eqs.~(\ref{eq:n_fer}), 
(\ref{eq:P_fer}) and (\ref{eq:rhoWD}) provides the basic EoS of 
electron-pressure dominated high-density condensed matter in this regime. 

There exist opposite limits to the cold fermion gas equation of state: 
the low density, nonrelativistic ($x\ll 1$) and the high density, 
extremely relativistic ($x\gg 1$) limits. 
From Eqs.~(\ref{eq:n_fer}) and (\ref{eq:rhoWD}) we find that in cold matter 
with $Y_e=0.5$ the electron relativity parameter satisfies $x_e=1$ 
at $\rho\approx 10^6\;$\gmcmc, which therefore marks the transition density 
between a nonrelativistic (NR) and extremely relativistic (ER) electron gas. 
It is convenient to write the EoS in the two limits in a 
{\it polytropic} form,
\begin{equation}\label{eq:polytrope}
P(\rho)=K\rho^\Gamma\;
\end{equation}
where
\begin{eqnarray}
\mbox{NR}  & x_e\ll 1,\; \rho\ll 10^6
\mbox{gm}\;\mbox{cm}^{-3}\;\;: 
& \Gamma=\frac{5}{3},\; K=1.0036 \times 10^{13} Y_e^{5/3} \;;
\label{eq:polyEoSNR}\\
\mbox{ER} & x_e\gg 1,\; \rho\gg 10^6
\mbox{gm}\;\mbox{cm}^{-3}\;\;:
& \Gamma=\frac{4}{3},\; K=1.2435 \times 10^{15} Y_e^{4/3} 
\label{eq:polyEoSER}\;.
\end{eqnarray}
The constant $K$ is in cgs units, yielding a pressure 
in \dncms for a density in \gmcmc. Note that in this approximation the 
composition of the matter enters only through $Y_e$. Correspondingly, 
helium, carbon and oxygen, which all have $Y_e=0.5$ have identical ideal 
equations of state, slightly stiffer than that of iron 
($Y_e=26/56\approx0.43$). Fully equilibrated matter (often referred to as 
``catalyzed'') has a $Y_e$ that decreases with density and is therefore softer 
than matter composed of a single element.

The free, degenerate electron pressure EoS outlined here is a 
good approximation for the equation of state below neutron-drip density. 
It was employed by Chandrasekar in his pioneering analysis of 
equilibrium white dwarfs \cite{ChandraWD}, 
for which he received the Nobel prize in 1983. More exact treatments were 
introduced in later years, which included the two main required 
corrections - electrostatic effects at low densities, and neutronization 
(or inverse $\beta-$decay) at higher densities. 

\subsubsection{Electrostatic Corrections to the Cold Equation of State: 
$\rho\lesssim10^4\;$\gmcmc}
\label{subsubsect:Electro} 

There exists a net electrostatic correction to the ideal equation of state 
due to the fact that the local distribution of charge is very nonuniform. 
The fact that positive charge is concentrated in point-like 
ions causes the average electron-ion separation to be smaller than 
the average distance between electrons. The net electrostatic potential 
felt by the electrons is thus an attractive one, which effectively reduces 
the pressure for a given density.

Electrostatic corrections to the cold equation state are mostly important at 
relatively low densities. Electrostatic energies are inversely 
proportional to the average separation between particles, $<\!r\!>$ 
which is naturally proportional to $n_e^{-1/3}$. The relative importance of 
electrostatic energy, $E_C$, between a degenerate, nonrelativistic electron 
and an ion of charge $Z$, can be estimated through 
\begin{equation}\label{E_CE_F}
\frac{E_C}{E_F^\prime}\equiv\frac{Ze^2/<r>}{p_{F,e}^2/2m_e}\propto n_e^{-1/3}
\end{equation}
where $E^\prime_{F,e}=p_{F,e}^2\propto n_e^{2/3}$ is the Fermi kinetic 
energy of the nonrelativistic electrons. 
Unlike the case of hot matter (where the mean electron kinetic energy is 
$\sim k_B T$), the relative importance of electrostatic 
corrections {\it decreases} with density as $n_e^{-1/3}$. 

A rough estimate of the quantitative electrostatic correction can be performed
by using the Wigner-Seitz approximation, which describes the lattice as 
neutral sphere with a central point-like ion and an ambient  uniform 
electron gas\footnote{We note that such an approximation is actually 
better suited for white dwarfs than for laboratory solids, where the electron 
distribution is much more nonuniform.}. For a carbon or oxygen 
lattice, the correction to the pressure is typically of the order of a few 
percent. At lower densities the electron distribution 
deviates from uniformity, and more sophisticated approaches, namely 
the {\it Thomas-Fermi} and {\it Thomas-Fermi-Dirac} models, must be invoked 
\cite{Sal61, FMT}. 

\subsubsection{Neutronization Corrections to the Cold Equation of State: 
$10^9\;$\gmcmc$\rho\lesssim 4\times 10^{11}\;$\gmcmc}
\label{subsubsect:neutron} 

Stable high density matter must be in chemical equilibrium to all types of 
reactions, including the weak interactions which drive $\beta$ 
decay and electron capture (``inverse $\beta-$decay''):
\begin{equation}\label{eq:betas}
n\rightarrow p+e+\bar{\nu}_e \;\;\;,\;\;\; p+e\rightarrow n+\nu_e\;,
\end{equation}
where $n$ and $p$ denote a neutron and a proton, respectively, $e$ denotes 
and electron and $\nu_e (\bar{\nu}_e)$ denote an electron neutrino 
(anti-neutrino). If the matter's composition is out of $\beta-$equilibrium, 
it will adjust through $\beta-$decays or electron capture. Both types of 
reactions change the electron per nucleon fraction, $Y_e$, and thus 
affect the EoS (Eqs.~(\ref{eq:polyEoSNR}-\ref{eq:polyEoSER})).

In cold white dwarfs and neutron stars the weakly interacting neutrinos 
freely escape the system: a zero neutrino abundance implies a zero neutrino 
chemical potential. The condition of chemical equilibrium is then stated as:
\begin{equation}\label{eq:mueq}
\mu_n=\mu_p + \mu_e\;,
\end{equation}
where $\mu_x$ denotes the chemical potential of species $x$.
The condition of chemical equilibrium is the fundamental origin of the stable 
existence of neutrons in nuclei and in uniform $n-p-e$ matter. For free, 
single, particles the chemical potential is identical to the rest mass. 
The masses in MeV of the three elementary particles of 
Eq.~(\ref{eq:mueq}) are $m_n=939.6,\;m_p=938.3$ and $m_e=0.511$ 
(1 MeV is equivalent to $\sim 1.78\times10^{-27}\;$gm). It is 
energetically allowed to have a {\it free} neutron decay in the reaction 
$n\!\rightarrow\!p\!+\!e\!+\!\bar{\nu}_e$ ($m_n-m_p-m_e\approx 0.8\;
\mbox{MeV}$). Indeed, the life-time of a free neutron is 
only about $1000\;$sec before it undergoes a $\beta$-decay. 
By contrast, in a cold, degenerate, noninteracting $n-p-e$ gas 
neutron decay can be ``blocked'': since the protons and electrons must obey 
the Pauli principle, the decays will be suppressed if the energy available to 
the newly formed electron and proton is insufficient to place them 
above their respective Fermi levels. At such high densities, the equilibrium 
state of the gas includes a finite fraction of neutrons. 

In bulk matter, the electron chemical potential, which is equal to the 
electron Fermi energy, rises with electron number density. Hence, maintaining 
chemical equilibrium (Eq.~\ref{eq:mueq})) may require some protons to capture 
electrons and covert to neutrons. In matter below the neutron drip density 
these conversions occur in nuclei, so their neutron fraction increases - 
they ``neutronize''. The result is a net decrease in the 
electron abundance ($Y_e$) at high densities, which lowers the pressure 
and ``softens'' the EoS. 

The fact that ``neutronization'' becomes energetically favorable at densities 
below the neutron drip must be taken into account when formulating an exact 
EoS for this range. In matter composed of nuclei and free electrons, 
$\beta-$decay is limited to relatively high densities 
($\rho\geq 10^9\;$\gmcmc), and the exact threshold 
for the onset of neutronization depends on the nature of the nucleus, since 
the nuclear interactions must be taken into account in addition 
to Fermi kinetic energies. For example, the reaction 
$^{12}_6 C + e \rightarrow ^{12}_5B$ 
requires the nucleus to absorb an energy of $\sim 14\;$MeV, which an 
electron can supply if the density of a pure 
carbon lattice reaches $\sim3\times 10^{10}\;$\gmcmc. The EoS  
of matter close to the neutron drip density is thus dependent on its chemical 
composition even for elements of equal $Y_e$, such as helium 
and carbon \cite{Sal61}.

\subsection{The Equation of State Above Neutron Drip Density: 
$\rho\gtrsim 4\times 10^{11}\;$\gmcmc}

As neutronization proceeds, the nuclei become increasingly neutron-rich. 
The so-called ``tensor-force'' component of the nuclear interactions causes 
like nucleons to repel one another, so that the binding energy of a neutron 
rich nucleus is smaller than in one where $Z/A=0.5$. Fully equilibrated 
matter, reaches the last stable isotope $^{118}$Kr ($Y_e\approx0.31$) 
just before the density of 
$\rho\equiv\rho_{n-drip}\approx 4.3\times 10^{11}\;$\gmcmc. 
At higher densities it becomes favorable for some neutrons to ``drip'' 
out of their parent nuclei. The nucleon component of the matter can no 
longer be confined to point-like objects and the 
nuclei begin to dissolve. As density is increased, 
a larger fraction of the nucleons exists as ``free'' particles, outside 
the nuclei. This is a gradual transition, until the 
matter approaches the density of atomic nuclei, where all nuclei have 
essentially dissolved, and the distribution of nucleons becomes uniform. 
The EoS above the neutron drip density must be formulated by consistently 
including the effects of nuclear physics, which become the governing 
component of the properties of matter as the density of 
atomic nuclei, $\rho_{nuc}\!\approx\!2.8\!\times\!10^{14}\;\mbox{gm/cm}^3$ 
is approached. 

\subsection{Subnuclear Densities: 
$4\times 10^{11}\;$\gmcmc$\lesssim\rho\lesssim 2.8\times 10^{14}\;$\gmcmc}

An analysis of cold matter in the range 
$\rho_{n-drip}\leq\rho\leq\rho_{nuc}$ is complicated by requiring chemical 
equilibrium between nucleons inside the nuclei and those that have 
dripped outside. 
One must account for the effects of the surrounding gas of free nucleons 
on the nuclei, as well as other effects such nuclei surface and Coulomb 
energies. In addition, the nuclei are expected to be very neutron rich, 
deviating from the $Z/A=0.4-0.5$ found in terrestrial nuclei. 
Nonetheless, the properties of matter in this range of densities can still be 
derived by a natural extrapolation from ordinary nuclei, and, indeed, the 
EoS for such matter is believed to be well understood. The principal studies 
of matter (e.g., Ref.~\cite{BBP}) are based on the nuclear liquid-drop 
model, and are suitable for most applications regarding neutron star 
structure. For problems where a more accurate description of the neutron 
star inner crust is required, attention must be given to 
lattice effects, as the nuclei and free nucleons arrange in a distinct 
spatial structure (where the nuclei settle into bubbles, slabs or rods, 
depending on density \cite{RavPeth}). 
It is expected that at $\rho\approx\frac{1}{2}\rho_{nuc}$ all nuclei will 
have dissolved so that the matter is completely uniform. 

\subsection{Supernuclear Densities: $\rho\gtrsim 2.8\times 10^{14}\;$\gmcmc}
        
The upper end of the high density regime is referred to as ``supernuclear'' 
where $\rho\geq\rho_{nuc}$. As matter is compressed to such densities the 
EoS becomes gradually dominated by the degenerate nucleons and the 
nucleon interactions.  

It is instructive to begin by considering an ideal uniform mixture 
of neutrons, protons and electrons. For any given nucleon number density, 
$n_N$, solving the equilibrium composition $n_n,n_p$ and $n_e$ 
(the neutron, proton and electron number densities, respectively) requires  
three equations. The first is baryon number conservation,
\begin{equation}\label{eq:n_nucleon}
n_N=n_n+n_p\;,
\end{equation}
and the other two arise from imposing chemical equilibrium 
(Eq.~\ref{eq:mueq}) and charge neutrality. 
The condition for charge neutrality is
\begin{equation}\label{eq:charg0}
n_p=n_e\;.
\end{equation}

For a noninteracting mixture of nucleons and electrons 
all chemical potentials are simply the Fermi energies (Eq.\ref{eq:E_F})). 
Electrons are extremely relativistic at nuclear densities 
so their electron chemical potential is 
\begin{equation}\label{eq:mue_xe-rhoB}
\mu_e\approx p_{F,e}c = \hbar c(3\pi^2 Y_e \rho_B)^{1/3}\approx 
 100\left(\frac{Y_e}{0.03}\right)^{1/3}\left(\frac{n_N}{n_{nuc}}\right)^{1/3}
   \mbox{MeV} \;\;,
\end{equation}
where $n_{nuc}=0.16\;\mbox{fm}^{-3}$ is the number density of nuclei at the 
saturation density (and $1\;\mbox{fm}\!=\!10^{-13}\;$cm). 
Even if the electron 
fraction per baryon is only a few percent, the electron chemical potential 
exceeds the mass difference between neutrons and protons by two orders of 
magnitude. The only way to maintain a finite electron fraction (which is 
required to balance the protons for charge neutrality) {\it and} 
satisfy chemical equilibrium is by having a significantly 
larger neutron than proton fraction (note that for nonrelativistic fermions 
$E_F\propto n^{2/3}$ and for extremely relativistic fermions 
$E_F\propto n^{1/3}$). Equilibrium matter at nuclear densities must be very 
neutron dominated, and objects composed of such matter are thus 
``neutron stars''. For an noninteracting gas the ratio of neutrons to 
protons in equilibrium must be about $8\!:\!1$, which is also 
representative of more realistic models of supernuclear densities. 

The noninteracting gas approximation is not reliable for deriving the EoS 
at supernuclear densities. Unlike electrostatic perturbations, 
nucleon-nucleon interactions are not negligible, and the interaction energies 
are comparable to the Fermi energies of the degenerate nucleons 
(electrons do not feel the strong interaction, and may still be treated as 
noninteracting). Modeling of the nucleon-nucleon interaction is one of 
the longest-standing problems in nuclear physics, still only partially 
solved. Profound difficulties exist due to the absence of a comprehensive 
theory of the interactions and the difficulty of obtaining experimental data 
for $\rho>\rho_{nuc}$. Further complications arise due to the fact 
the Fermi and interaction energies at $\rho\sim2-3\rho_{nuc}$ reach a sizable 
fraction of the rest mass, and relativistic effects must be taken into 
account as well. It is not yet possible to apply quantum chromodynamics 
(QCD), the fundamental theory of strong interactions, to the many-body 
nuclear domain at $\rho\approx\rho_{nuc}$. Instead, the most useful 
approaches are still based 
on phenomenological potential formalisms, and many-body Schr\"odinger-like 
systems of equations \cite{ShaTeu83}. 

The EoS of supernuclear matter remains to date a field of active research. 
Current approaches include variational 
methods based on deduced two and three nucleon interactions (\cite{WFF}, 
see {\cite{APR} for a recent study), and relativistic mean field 
approximations (\cite{SerWal79}, see \cite{Glenbook} for a review). 
The models are based on fitting parameters to reproduce the empirically 
determined properties of finite nuclei with $n=n_{nuc}$. A rough 
approximation of the properties of the nucleon component of supernuclear 
matter can be obtained through the effective, nonrelativistic model of 
Ref.~\cite{LatSwes},  
\begin{equation}\label{eq:LSmodel}
\rho(n,x)=E_{gas}(n,x)+a n^2 +b (n_n-n_p)^2 +c n^{\delta+1}\;.
\end{equation}
The first term $E_{gas}$ is the mass-energy density of 
a noninteracting gas of nucleons at density $n$ and composition $n_n,n_p$. 
The coefficient $a$ is negative, representing the long range attractive 
component of the inter-nucleon force, while $c$ is positive, representing 
the short-range repulsive component. The power $\delta$ is larger than unity, 
so that short-range repulsion dominates at high density. The {\it symmetry} 
term includes the positive coefficient $b$ which describes the 
``tensor-force'' that repels like nucleons, and its contribution is therefore 
minimized at for symmetric nuclear matter ($n_n=n_p$). The values 
of $a,b,c$ and $\delta$ are derived by requiring Eq.~(\ref{eq:LSmodel}) to 
reproduce the assumed properties of symmetric matter at $n_N=n_{nuc}$. 
See Refs.~\cite{LatSwes,BG} for typical (model dependent) values of these 
parameters. 

Finally, we note that it is quite possible that other particles, besides 
neutrons, protons and electrons, coexist 
in stable equilibrium at supernuclear matter. The most obvious example is 
the muon, which is a lepton similar to the electron, but has a rest mass of 
$m_\mu\approx 105\;\mbox{MeV}$. The equilibrium condition for the muons 
will simply be $\mu_\mu=\mu_e$, so from Eq.~(\ref{eq:mue_xe-rhoB}) it is 
evident that at densities $n\gtrsim n_{nuc}$ it is energetically favorable to 
convert some electrons into muons through the weak interactions. At 
densities of $n\gtrsim 2 n_{nuc}$ it possible that other exotic particles 
will appear, such as hyperons (baryons that are heavier than nucleons), 
Bose-Einstein condensates of mesons (i.e., pions or kaons), or even 
conversion of the nucleons into an uniform mixture of quarks 
(see Refs.~\cite{Glenbook, Prakal97, BLC} for recent reviews of the possible 
presence and roles of such particles in neutron stars). 

\subsection{Basic Properties of High Density Matter}

In Table \ref{tab:EoS} we present an 
EoS of cold, fully catalyzed matter ranging from normal 
iron at zero pressure to supernuclear densities. The EoS in the subnuclear 
regime is compiled from Refs.~\cite{BPS,FMT,BBP} 
and are ``standard'' for studying catalyzed high density matter. For 
realistic white dwarf models, which are presumably composed mostly of helium 
or carbon and oxygen, and never reached sufficiently high temperatures to 
catalyze their nuclei, a slightly different EoS must be used, based on a 
single species; see Ref.~\cite{Sal61}. We also tabulate one 
``state-of-the-art'' EoS for 
the supernuclear range \cite{APR} 

\begin{table*}
\caption{\label{tab:EoS} The equation of state $^a$ of cold, catalyzed 
high density matter$^b$}
\begin{center}
\footnotesize
\begin{tabular}{c c c c c c c}
$\rho$   &   $n_B$   &   $P$    & $\Gamma$  &   $Y$  &   $K$   & $c_s$ \\ 
\gmcmc   & cm$^{-3}$ & \dncms   &           & \dncms &   MeV  & \cms  \\ 
\hline
1.00E+01 & 6.02E+24 & 5.13E+11 & 7.03E+00 & 3.20E+12 & 7.68E-06 & 5.67E+05 \\
2.00E+01 & 1.21E+25 & 2.59E+13 & 3.09E+00 & 7.32E+13 & 3.41E-05 & 1.91E+06 \\
5.00E+01 & 3.01E+25 & 2.21E+14 & 2.50E+00 & 5.51E+14 & 1.03E-04 & 3.32E+06 \\
1.00E+02 & 6.02E+25 & 1.13E+15 & 2.23E+00 & 2.88E+15 & 2.69E-04 & 5.37E+06 \\
2.00E+02 & 1.20E+26 & 5.14E+15 & 2.14E+00 & 1.10E+16 & 5.15E-04 & 7.43E+06 \\
5.00E+02 & 3.01E+26 & 3.54E+16 & 2.05E+00 & 7.27E+16 & 1.36E-03 & 1.21E+07 \\
1.00E+03 & 6.02E+26 & 1.44E+17 & 1.90E+00 & 2.95E+17 & 2.75E-03 & 1.72E+07 \\
2.00E+03 & 1.20E+27 & 4.91E+17 & 1.74E+00 & 8.56E+17 & 3.99E-03 & 2.07E+07 \\
5.00E+03 & 3.01E+27 & 2.53E+18 & 1.82E+00 & 4.61E+18 & 8.61E-03 & 3.04E+07 \\
1.00E+04 & 6.02E+27 & 9.01E+18 & 1.80E+00 & 1.75E+19 & 1.63E-02 & 4.18E+07 \\
2.00E+04 & 1.20E+28 & 3.07E+19 & 1.78E+00 & 5.45E+19 & 2.54E-02 & 5.22E+07 \\
5.00E+04 & 3.01E+28 & 1.51E+20 & 1.73E+00 & 2.61E+20 & 4.87E-02 & 7.22E+07 \\
1.00E+05 & 6.02E+28 & 4.90E+20 & 1.69E+00 & 8.91E+20 & 8.32E-02 & 9.44E+07 \\
2.00E+05 & 1.20E+29 & 1.57E+21 & 1.65E+00 & 2.59E+21 & 1.21E-01 & 1.14E+08 \\
5.00E+05 & 3.01E+29 & 7.07E+21 & 1.60E+00 & 1.13E+22 & 2.11E-01 & 1.50E+08 \\
1.00E+06 & 6.02E+29 & 2.16E+22 & 1.59E+00 & 3.63E+22 & 3.39E-01 & 1.90E+08 \\
2.00E+06 & 1.20E+30 & 6.38E+22 & 1.57E+00 & 9.97E+22 & 4.65E-01 & 2.23E+08 \\
5.00E+06 & 3.01E+30 & 2.63E+23 & 1.48E+00 & 3.90E+23 & 7.29E-01 & 2.79E+08 \\
1.00E+07 & 6.02E+30 & 6.81E+23 & 1.44E+00 & 1.03E+24 & 9.64E-01 & 3.21E+08 \\
2.00E+07 & 1.20E+31 & 1.88E+24 & 1.44E+00 & 2.68E+24 & 1.25E+00 & 3.66E+08 \\
5.00E+07 & 3.01E+31 & 6.84E+24 & 1.40E+00 & 9.55E+24 & 1.78E+00 & 4.37E+08 \\
1.00E+08 & 6.02E+31 & 1.78E+25 & 1.38E+00 & 2.57E+25 & 2.40E+00 & 5.07E+08 \\
2.00E+08 & 1.20E+32 & 4.67E+25 & 1.36E+00 & 6.29E+25 & 2.94E+00 & 5.61E+08 \\
5.00E+08 & 3.01E+32 & 1.53E+26 & 1.39E+00 & 2.11E+26 & 3.95E+00 & 6.50E+08 \\
1.00E+09 & 6.02E+32 & 3.91E+26 & 1.17E+00 & 4.57E+26 & 4.27E+00 & 6.76E+08 \\
2.00E+09 & 1.20E+33 & 8.83E+26 & 1.36E+00 & 1.19E+27 & 5.57E+00 & 7.72E+08 \\
5.00E+09 & 3.01E+33 & 3.03E+27 & 1.36E+00 & 4.12E+27 & 7.71E+00 & 9.07E+08 \\
1.00E+10 & 6.01E+33 & 7.24E+27 & 1.35E+00 & 1.01E+28 & 9.47E+00 & 1.01E+09 \\
2.00E+10 & 1.20E+34 & 1.83E+28 & 1.24E+00 & 2.24E+28 & 1.05E+01 & 1.06E+09 \\
5.00E+10 & 3.00E+34 & 5.63E+28 & 1.13E+00 & 6.37E+28 & 1.19E+01 & 1.13E+09 \\
1.00E+11 & 5.99E+34 & 1.40E+29 & 1.26E+00 & 1.80E+29 & 1.69E+01 & 1.34E+09 \\
2.00E+11 & 1.20E+35 & 3.13E+29 & 1.17E+00 & 3.69E+29 & 1.73E+01 & 1.36E+09 \\
5.00E+11 & 2.99E+35 & 8.28E+29 & 5.60E-01 & 4.65E+29 & 8.75E+00 & 9.63E+08 \\
1.00E+12 & 5.97E+35 & 1.26E+30 & 8.48E-01 & 1.05E+30 & 9.86E+00 & 1.02E+09 \\
2.00E+12 & 1.19E+36 & 2.12E+30 & 7.53E-01 & 1.61E+30 & 7.60E+00 & 8.97E+08 \\
5.00E+12 & 2.98E+36 & 4.79E+30 & 1.11E+00 & 5.34E+30 & 1.01E+01 & 1.03E+09 \\
1.00E+13 & 5.95E+36 & 1.16E+31 & 1.44E+00 & 1.76E+31 & 1.66E+01 & 1.33E+09 \\
2.00E+13 & 1.19E+37 & 3.26E+31 & 1.52E+00 & 4.94E+31 & 2.34E+01 & 1.57E+09 \\
5.00E+13 & 2.97E+37 & 1.32E+32 & 1.53E+00 & 2.03E+32 & 3.84E+01 & 2.01E+09 \\
1.00E+14 & 5.92E+37 & 3.93E+32 & 1.87E+00 & 8.36E+32 & 7.94E+01 & 2.88E+09 \\
2.00E+14 & 1.18E+38 & 1.93E+33 & 2.75E+00 & 5.54E+33 & 2.64E+02 & 5.24E+09 \\
5.00E+14 & 2.89E+38 & 1.77E+34 & 3.00E+00 & 5.54E+34 & 1.08E+03 & 1.03E+10 \\
1.00E+15 & 5.50E+38 & 1.52E+35 & 2.87E+00 & 6.52E+35 & 6.68E+03 & 2.30E+10 \\
2.00E+15 & 9.36E+38 & 9.20E+35 & 2.23E+00 & 3.01E+36 & 1.81E+04 & $^c$     \\ 
5.00E+15 & 1.57E+39 & 4.87E+36 & 1.50E+00 & 1.44E+37 & 5.14E+04 & $^c$     \\ 
\hline
\end{tabular}
\\
\end{center}
\renewcommand{\baselinestretch}{1.0}

\vspace{-0.2cm}
$\;\;^a$ - 
{\scriptsize shortened notation: $2.00E+10$ is $2.00\times10^{10}$}.\\
\vspace{-0.2cm}
$^b$ - 
{\scriptsize The EoS up to $\rho\sim\rho_{nuc}=2.8\times 10^{14}$ is 
based on Table 5 of 
Ref.~\cite{BBP} [$\copyright$ 1971. The American Astronomical Society].
The supernuclear EoS is adapted from Ref.~\cite{APR}}.\\
\vspace{-0.2cm}
$^c$ - 
{\scriptsize the APR \cite{APR} EoS is not relativistic, and at extremely 
high densities has an unphysical superluminal (greater than light speed) 
speed of sound.}
\end{table*}

\begin{figure}[htb]
\begin{center}
\includegraphics[width=10cm]{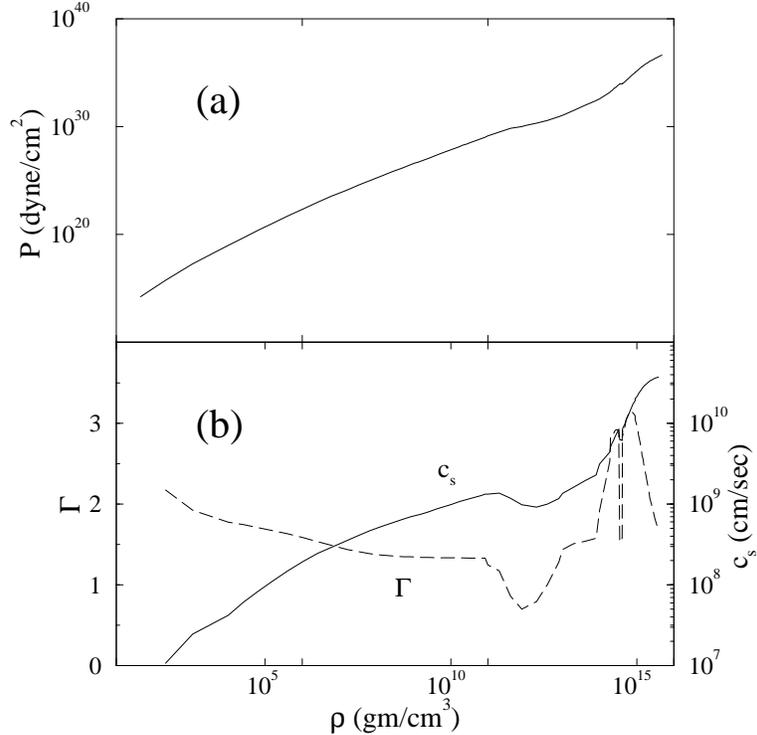}
\caption{The equation of state of cold, catalyzed high density matter, 
based on Refs.~\cite{FMT, BBP, BPS, APR}: (a) the pressure density relation, 
(b) adiabatic index and sound speed. \label{fig:EoS}}
\end{center}
\end{figure}

The EoS $P(\rho)$ is plotted in Figure \ref{fig:EoS}a. It is 
especially instructive to examine corresponding values of the adiabatic 
index, $\Gamma\equiv d \ln\!P/d \ln\!\rho$, plotted in Figure \ref{fig:EoS}b, 
along with the sound speed ($c_s=\sqrt{d\!P/d\!\rho}$).  
Note that around $\rho\approx 10^6\;$\gmcmc~ the EoS transforms, as expected, 
from a nonrelativistic $\Gamma=5/3$ polytrope to a relativistic 
$\Gamma=4/3$ one as the Fermi energy of the electrons, which 
dominate the pressure, gradually becomes relativistic. There is a sharp drop 
in the adiabatic index around the neutron drip density, since at first 
the dripped neutrons contribute to mass-density but not to pressure. Only as 
matter approaches nuclear densities does the nucleon pressure become 
important, pushing the adiabatic index back up to values in the range $2-3$.  
 
Along with the EoS, we also examine some thermodynamic quantities of 
cold high density matter. In Table \ref{tab:EoS} we also list 
the adiabatic index, sound speed, 
bulk modulus $Y$, and incompressibility, $K$, defined as
\begin{equation}\label{eq:YandK}
Y=n \frac{dP}{dn}\;\;\; \mbox{and} \;\;\; 
K=9\frac{dP}{dn}\equiv 9n\frac{d^2\rho}{dn^2}\;. 
\end{equation}
Note that the bulk modulus (which has 
units of pressure) is the reciprocal of the quantity usually defined as 
the compressibility ($\chi\equiv Y^{-1}$), while the ``incompressibility'' 
(units of energy) is more commonly used in nuclear physics applications 
(and is indeed measurable in nuclei).  
In Table \ref{tab:moduli} we list typical values of $\Gamma,\;c_s,\;Y,\;K$ 
and the specific heat capacity, $c_v$, for condensed matter 
found in white dwarfs and neutron stars. For the purpose of comparison, we 
also give values for the sun, where the pressure roughly follows an 
ideal Maxwell-Boltzmann law $P\sim n k_B T$. Note how the extreme 
conditions of high densities leads to properties that are very 
different than those found for terrestrial materials. For example, the speed 
of sound in a neutron star is expected to be several tens of percent of the 
speed of light!     

\begin{table}[h] 
\caption{Typical values of thermodynamic quantities in the sun, white dwarfs 
and neutron stars
.\label{tab:moduli}}
\begin{center}
\footnotesize
\begin{tabular}{l c c c c c}
Object &  $\Gamma$   &   $c_s$    &     $Y$    &     K         
&   $c_v$   \\
       &             &   (\cms)   & (\dncms)   &    (MeV)   
& \ergsdeggm   \\ \hline
Sun &         
    $5/3$ &       $\sim10^7$     & $\sim10^{12}$ &  $\sim 10^{-4}$  
& $\sim\frac{3}{2}\left(1/\mu+1/m_p\right)k_B$   \\
White Dwarf (C/O) & 
    $4/3$ & $\sim 5\times 10^8$  & $\sim10^{24}$ & $\sim 10$ 
& $\frac{3}{2}k_B/\mu$   \\
Neutron Star &
    $2-3$   & $\sim10^{10}$  & $\sim10^{35}$ & $\sim 200 $
& $\sim 10^{-3}k_B/m_p\;^\dagger$ \\ \hline
\end{tabular}
\\
{\footnotesize $\dagger$ - 
for an ideal nonrelativistic neutron gas and a central 
temperature of $\sim 10^9\;$\Kelv $\rightarrow k_B T/E_F\approx 10^{-3}$.}
\end{center}
\end{table}

\section{White Dwarf and Neutron Star Structure}
\label{sect:WDNSconf}
 
Compact objects are self-gravitating equilibria with a mass comparable to 
that of the sun, $1\;M_{\odot}\equiv 1.989\times 10^{33}\;$grams. 
Both white dwarfs and neutron stars are {\it centrally condensed} objects - 
most of their mass is located in a high density core, which is limited to 
a fraction of the volume; furthermore, the radius of the objects 
{\it decreases} with increasing mass. These traits are characteristic of a 
configuration supported by degenerate-fermion pressure.

\subsection{Construction of a Self-gravitating, Equilibrium Star}
\label{subsect:HydStat}

The equilibrium structure of a self-gravitating object is derived from the 
equations of hydrostatic equilibrium. The simplest case is that of a 
spherical, nonrotating, static configuration, where for a given EoS all 
macroscopic properties are parameterized by a single parameter, for example, 
the central density. In the case of compact objects, the gravitational 
fields are strong enough that calculations must be performed in the context 
of general relativistic (rather than Newtonian) 
gravity. The fundamental equation of hydrostatic equilibrium in its 
general relativistic form has been 
derived by Tolman \cite{Tolman} and Oppenheimer and Volkoff \cite{OV}, 
and is known as the ``TOV'' equation:   
\begin{equation} \label{eq:TOV}
\frac{dP(r)}{dr}=-\frac{Gm(r)\rho(r)}{r^2}
               \left(1+\frac{P(r)}{c^2\rho(r)}\right)
               \left(1+\frac{4\pi r^3P(r)}{c^2m(r)}\right)
                       \left(1-\frac{2Gm(r)}{c^2r}\right)^{-1} \;\;.
\end{equation}
This equation simply states that at any radial distance $r$, the 
gravitational pull by the mass interior to $r$, $m(r)$, is balanced by the 
gradient of the pressure $P(r)$; 
$G=6.67\times 10^{-8}\;\mbox{cm}^3\;\mbox{gm}^{-1}\mbox{sec}^{-2}$ is the 
gravitational constant. Note that the first term on the right hand side is 
the only term in the nonrelativistic case, 
where $P/(\rho c^2)\ll 1$ and $2Gm/(rc^2)\ll1$. The second and third factors 
arise from the pressure being a form of energy density (``regeneration of 
pressure'' effect), while the last term includes the correction due to the 
curvature of space in the strong gravitational field of the star.

\subsection{Stable Configurations of High Density Matter}
\label{subsect:Confs}

Given an EoS, Eq.~(\ref{eq:TOV}) 
can be integrated simultaneously with the mass equation, 
\begin{equation} \label{eq:dmdr}
   \frac{dm(r)}{dr}=4\pi r^2\rho(r)\;\;,
\end{equation}
to determine the entire profile of the object. The actual integration is 
typically solved by numerical means, assuming a central density 
$\rho(r=0)\!=\!\rho_c$, and integrating outward from the center until 
reaching the surface where $P(r)\!=\!0$, which identifies $r=R$ as the radius 
of the star and $M(R)$ as its mass. We emphasize that the TOV equations 
provide the {\it gravitational} mass (or total mass-energy) of 
the star, which includes the effect of the gravitational binding energy, 
$E_{GB}\sim G M^2/R$, as well as the internal and rest mass energy of 
the stellar constituents. Note that in the case of neutron stars, 
$(G M^2/R)/(M c^2)\sim 10-20\%$; indeed, the gravitational mass of a neutron 
star is measurably lower than the total rest mass of its constituents.

\begin{figure}[htb]
\begin{center}
\includegraphics[width=10cm]{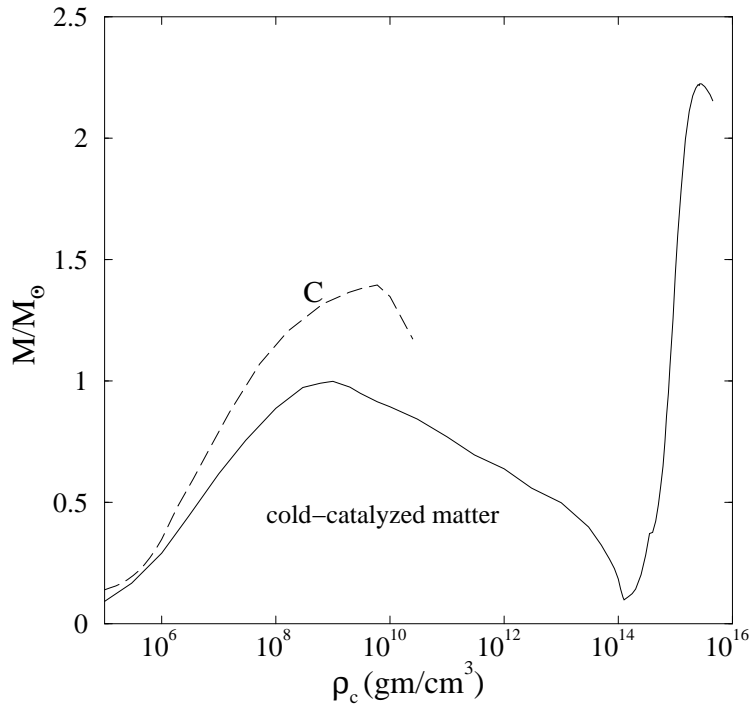}
\caption{Mass vs.~central density of a cold, carbon star \cite{HamSal61} 
(dashed line) and of a star composed of cold catalyzed matter 
(solid line). \label{fig:M_vs_rhoc}}
\end{center}
\end{figure}

Varying $\rho_c$ produces a {\it sequence} of models for the 
given EoS, yielding $M(\rho_c)$ and $R(\rho_c)$ along the sequence. 
The resulting sequences for stars composed of pure carbon 
(based on \cite{HamSal61}) and of cold, catalyzed condensed matter 
(with the EoS tabulated in table~\ref{tab:EoS}) are both shown in 
Figure \ref{fig:M_vs_rhoc} (note that carbon dominated matter beyond the 
neutron drip density is unphysical and therefore omitted). 
The most distinct feature about the equilibrium sequences is  
the existence of local maxima in the $M(\rho_c)$ curves. A hydrostatic 
equilibrium configuration is dynamically {\it unstable} to 
catastrophic gravitational collapse if $d M/d \rho_c < 0$, since a radial 
perturbation would cause it to collapse on itself \cite{ShaTeu83}. 
Therefore stable carbon white dwarfs exist only if the central density 
is $\rho_c\leq 6\times 10^9\;$\gmcmc, while cold, catalyzed, matter has 
two distinct regimes of stable configurations: one with central densities 
below $\sim 10^{10}\;$\gmcmc (equivalent to white dwarfs), and one where the 
central density lies roughly in the range 
$10^{13}\leq\rho_c\leq 10^{15}\;$\gmcmc (neutron stars). Configurations 
where $\rho_c\gtrsim \rho_{n-drip}$ are unstable, and cannot 
be found in nature. This situation is a direct consequence 
of the nature of the EoS as shown in Figure \ref{fig:EoS}. 
Quantitatively, it can be shown that a hydrostatic configuration is stable 
only if its mass averaged adiabatic index, $\bar{\Gamma}$, satisfies  
\cite{ShaTeu83}
\begin{equation}\label{eq:stabGR}
\bar{\Gamma}-\frac{4}{3}\gtrsim \frac{2GM}{Rc^2}\;. 
\end{equation}
It is evident therefore, that the effect of neutronization and neutron drip, 
which cause $\bar{\Gamma}$ to drop below $4/3$, is to 
separate astrophysical objects with high density matter into two 
distinct classes of stable configurations, which are shown schematically in 
Figure \ref{fig:Scheme}.

\begin{figure*}[t]
\begin{center}
\hspace{-2cm}
\includegraphics[angle=-90,width=11cm]{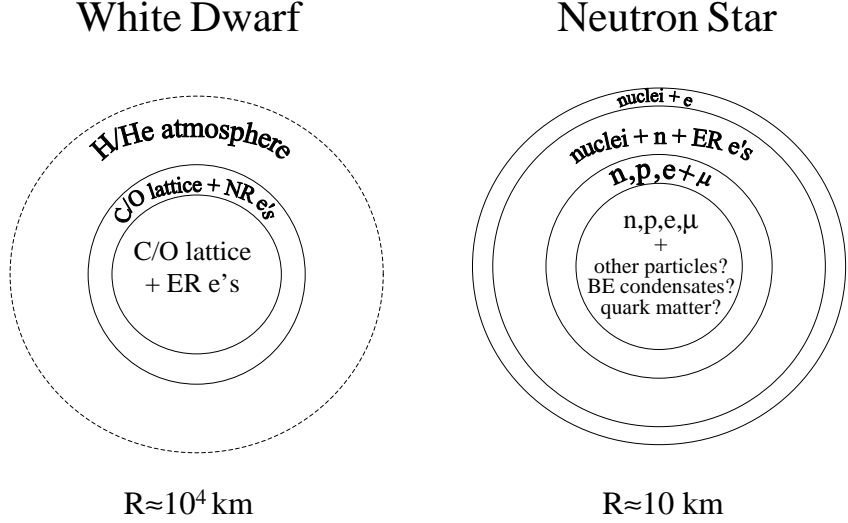}
\vspace{-5cm}
\caption{Schematic illustration of the structure of a typical white dwarf 
(left) and a neutron star (right). \label{fig:Scheme}} 
\end{center}
\end{figure*}

\subsubsection{White Dwarfs}

White dwarfs are composed of matter below the neutronization density. 
The astrophysical scenario which creates white dwarfs, 
i.e., quasi-static contraction of a progenitor star at the end point of 
thermonuclear burning, does not allow matter to reach high enough 
temperatures to achieve equilibrium composition through further burning. 
Accordingly, the composition of white dwarfs is mostly dominated by 
its nuclear ashes. In the standard scenario white dwarfs are composed mostly 
of carbon and oxygen, but there is observational evidence that helium 
white dwarfs (and perhaps even iron-core white dwarfs - see below) 
exist as well. The pressure throughout the star is due to degenerate 
electrons, which are nonrelativistic in the outer layers, but are 
relativistic in the interior of the more massive stars with $M\sim M_\odot$. 

By approximating the EoS as a Newtonian polytrope 
(Eq.~(\ref{eq:polytrope})),  
Chandrasekar \cite{ChandraWD} derived the basic features of white dwarfs:
\begin{eqnarray}\label{eq:WD-low}
\rho_c\lesssim 10^6\mbox{\gmcmc} & \Gamma=\frac{5}{3}:\;\;
&R\asim1.12\times 10^9\times
\left(\frac{\rho_c}{10^6\;\mbox{gm}\;\mbox{cm}^{-3}}\right)^{-1/6}
\left(\frac{Y_e}{0.5}\right)^{5/6} \;\mbox{cm} \\  
& & M\asim0.496\times 
\left(\frac{\rho_c}{10^6\;\mbox{gm}\;\mbox{cm}^{-3}}\right)^{1/2}
\left(\frac{Y_e}{0.5}\right)^{5/2} \;M_\odot\;,  \nonumber
\end{eqnarray}
\begin{eqnarray}\label{eq:WD-high}
\rho_c\gtrsim 10^6\mbox{\gmcmc} & \Gamma=\frac{4}{3}:\;\; & 
R\asim3.35\times 10^9\times 
\left(\frac{\rho_c}{10^6\;\mbox{gm}\;\mbox{cm}^{-3}}\right)^{-1/3}
\left(\frac{Y_e}{0.5}\right)^{2/3} \;\mbox{cm} \\
& & M\asim1.46\times 
\left(\frac{Y_e}{0.5}\right)^{2} \;M_\odot\;.  \nonumber
\end{eqnarray}
White dwarfs are expected to have radii of the order of $10^4\;$km - roughly 
the size of the Earth, or about one percent of that of the sun (which has 
a radius $R_\odot\approx 7\times 10^5\;$km). 

As long as the EoS is approximated as a pure polytrope, 
so that electrostatic and neutronization corrections are ignored, helium, 
carbon and oxygen white dwarfs  
($Y_e=0.5$) are all identical, while an iron-dominated white dwarf 
is only slightly different ($Y_e\approx 0.43$). Cold, catalyzed matter, 
which is used for the mass sequence in Figure \ref{fig:M_vs_rhoc}, has 
a significantly softer EoS (due to $Y_e$ decreasing with increasing density), 
and its sequence lies, therefore, lower than those of stars composed 
of a single species.  
The most significant result in Eq.~(\ref{eq:WD-high}) is that there is no 
dependence of $M$ on the central density. The mass 
$M_{Ch}\asim 1.46\;M_{\odot}$, known as the Chandrasekar mass, is the 
asymptotic value a white dwarf can reach if it achieves sufficiently high 
density, so that its entire 
structure is governed by relativistic fermions (in practice this mass cannot 
be reached, since the outermost layer have nonrelativistic electrons). 
Nonrotating white dwarfs cannot have a mass exceeding $M_{Ch}$, and the 
precise limit on their mass is lower by several percent, due to 
neutronization at high densities 
(\Sec~\ref{sect:WDNSconf}\ref{subsect:Confs}\ref{subsubsect:neutron}) 
{\it and} general-relativistic effects. In spite of the negligible effect on 
the hydrostatic equilibrium {\it profile} of white dwarfs, general relativity 
is required for a full analysis of white dwarf {\it stability}, 
since an exact $\Gamma=\frac{4}{3}$ star is unstable to gravitational 
collapse (Eq.~\ref{eq:stabGR}). 

\subsubsection{Neutron Stars}

Neutron stars are composed mostly of matter at nuclear 
densities, including a core with supernuclear densities, and is 
topped by a thin crust at subnuclear densities. The crust, 
composed of cold, catalyzed matter, may also be 
divided into an inner part with $\rho_{n-drip}\leq\rho\leq\rho_{nuc}$ 
and an outer part where $\rho\leq\rho_{n-drip}$.

Immediately after the neutron was discovered (1932) Landau modeled a neutron 
star as a gas of noninteracting, degenerate neutrons \cite{Landau32}. He 
found that neutron stars would have a maximum mass of about $1 M_\odot$ and 
a radius of several kilometers. Although this was a very crude 
approximation, it does suggest the correct orders of magnitude regarding 
the structure of these objects: a neutron star has a mass comparable to that 
of the sun compressed to the size of a medium city! If supported by 
degenerate fermion pressure, a neutron star cannot have a mass that exceeds 
the Chandrasekar limit by much, while its radius is inversely proportional to 
the mass of the pressure providing fermion. A neutron star should indeed have 
radius about $m_n/m_e\sim 10^3$ times smaller that of a white dwarf. 

\begin{figure*}[h]
\begin{center}
\includegraphics[width=10cm]{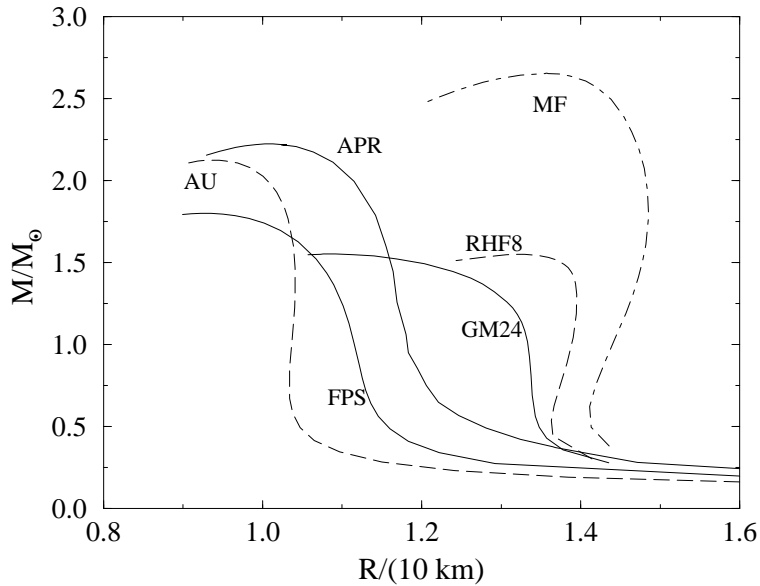}
\caption{\label{fig:MvsR}
The mass vs.~radius relation for of a nonrotating neutron star 
for several supernuclear equations of state: 
Huber et al.~(RHF8, \protect\cite{RHF8}), Pandharipande and Smith (MF, 
\protect\cite{MF}), Glendenning (GM24, \protect\cite{Glenbook}, page 244), 
Friedman and Pandharipande (FPS, \protect\cite{FPS}), 
Wiringa, Fiks, and Fabrocini (AU, \protect\cite{WFF}), and Akmal, 
Pandharipande and Ravenhall (APR, \protect\cite{APR}).}
\end{center}
\end{figure*}

A realistic treatment of neutron star structure requires general 
relativity \cite{OV}, whose effects are appreciable in this case. 
Furthermore, quantitative estimates must be based on a realistic EoS at all 
densities and especially a realistic model for the supernuclear regime 
where particle interactions are most important. In particular, 
these interactions are repulsive at short distances and oppose compression, 
thereby stiffening the EoS in comparison to a free particle gas. Indeed, 
plausible equations of state predict that $M_{max}(NS)\sim 2\;M_\odot$, 
($\sim 2.2\;M_\odot$ for Ref.~\cite{APR}), 
whereas the star's radius will lie in the range $R\sim 10-15\;$km. 
As an example, we show in Figure \ref{fig:MvsR} the $M$ vs.~$R$ relation 
found for a few representative equations of state of supernuclear densities.
The {\it exact} value of $M_{max}(NS)$ depends on the assumed EoS and 
therefore provides an integral measure of the properties of matter at 
supernuclear densities \cite{BLC, PPS76}. We also note that 
the existence of a maximum mass for neutron stars has important astrophysical 
implications, since it suggests that a larger mass cannot be sustained by 
cold pressure and must inevitably collapse to a black hole.   

\section{Observations of Condensed Matter in White Dwarfs}
\label{sect:WDs}

Over 2000 white dwarfs have been discovered to date, based on the 
spectroscopic properties of observed stars \cite{WDcataloge}. While the 
thermal energy in white 
dwarfs is small compared to the Fermi energies of the electrons in the 
interiors, it is still sufficient to generate an observable surface 
luminosity for several billion years. The luminosity of 
a star is dependent on its radius, $R$, and its effective 
surface temperature, $T_{eff}$, according to 
\begin{equation}\label{L_RT}
L=4 \pi \sigma R^2 T_{eff}^4\;,
\end{equation}
where $\sigma=5.67\times 10^{-5}\;
\mbox{erg}\;\mbox{cm}^{-2}\;\mbox{sec}^{-1}\;\mbox{deg}^{-4}$ is the 
Stephan-Boltzmann constant. The luminosities of white dwarfs are 
typically $10^{-3}$ to $10^{-2}$ of that of the sun, 
$L_\odot\approx 4\times 10^{33}\;$ergs sec$^{-1}$, but their radii are also
smaller than those of ordinary stars by a factor of $\sim 100$. 
As a result, white dwarfs have apparent surface temperature of 
several times $10^4\;$\Kelv, unusually high in comparison
with most observable stellar objects, so they do appear very ``white''. 
The existence of white dwarfs was established spectroscopically 
(by determining the surface temperature of sources) as early as 1910.

White dwarfs are believed to be remnants of stars with initial masses 
in the range $0.1-8\;M_\odot$, which are not 
massive enough to complete the thermonuclear burning process all the way to 
iron. Mostly they are mostly composed of carbon and oxygen, but in some 
cases, thermonuclear burning ceased before these elements were produced, and 
such stars are dominated be helium. 

\subsection{Radii} 

Observational determination of a white dwarf radius is straightforward 
if its flux, $F$, is measured and its distance from the Earth, $D$, is 
known: 
\begin{equation}\label{eq:RWD}
F(D)=\frac{L}{4 \pi D^2}\;\;\rightarrow \;\;R^2=
\frac{F D^2}{\sigma T_{eff}^4}\;.
\end{equation}
The stellar radius is derived only after the effective surface 
temperature is obtained spectroscopically \cite{Ship79}. 
Estimated radii of white dwarfs reside in the range 
of $0.007-0.013\;R_\odot$, which is consistent with an 
object supported by degenerate electron pressure 
(\Sec~\ref{sect:WDNSconf}\ref{subsect:Confs}) 
and thereby confirms the basic nature of white dwarfs. 

\subsection{Mass-Radius Relations}

The most significant test of the nature of matter in a white dwarf is 
obtained by comparing observed mass-radius relations with the theoretical 
predictions. Once the radius of a white dwarf is determined, details of the 
surface emission, namely effects of gravitational acceleration on line 
emission\cite{Schmidt96} (which scales as $M/R^2$), 
or gravitational redshift \cite{ShaTeu76} (which scales as $M/R$), 
are then used to estimate the surface gravity. 
An independent (and usually more accurate) estimate of the mass can be 
obtained directly for white dwarfs in binary-star systems, through 
Kepler's third law \cite{Greenstein86}. 
Through these methods, masses of a few white dwarfs have been 
estimated with reasonable accuracy, and most seem to cluster around 
$\sim 0.6\;M_\odot$ \cite{Weidermann90}. Some larger mass white dwarfs are 
also known, including the most famous Sirius B \cite{SiriusB}, with 
$M\approx 1\;M_\odot$. Note that these masses are smaller than 
$M_{Ch}$: it is believed that stellar evolution, and especially periods of 
mass loss, limit the masses of most white dwarfs to $M\leq 1\;M_\odot$, even 
though the EoS could sustain somewhat higher masses.

Some examples of observationally determined radii and mass for white dwarfs 
are presented in Table \ref{tab:obsWDs}. Note that larger mass stars have 
smaller radii, as is expected (Eqs.~(\ref{eq:WD-low}-\ref{eq:WD-high})). 
It is interesting to note that for some time, 
there was a nonnegligible discrepancy between the observations and 
theory, where observed radii seemed to be $\sim 10-20\%$ smaller than 
estimated by theory for carbon white dwarfs \cite{Schmidt96}. Only improved 
estimates of distances to several white dwarfs with the {\it Hipparcos} 
satellite, and better modeling of white dwarf atmospheres \cite{Wood95} 
have allowed this discrepancy to be mostly resolved \cite{Provetal98}. 
A comparison of current observed mass-radii determinations with theoretical 
curves is presented in Figure \ref{fig:obsWDs}, and in general, the fit is 
indeed very good: these new results seem to confirm that the 
composition of most white dwarfs is indeed dominated by medium weight 
elements (carbon and oxygen). However, they also imply that a small 
minority of white dwarfs do have relatively small radii, and may therefore 
contain iron cores, which 
presents an intriguing puzzle from the point of view of stellar evolution.  

\begin{table} 
\caption{Mass-radius relations of selected white dwarfs 
with $1\sigma$ errors. [Adapted from Provencal et al.~\cite{Provetal98}, 
$\copyright$ 1998. The American Astronomical Society] \label{tab:obsWDs}}
\begin{center}
\begin{tabular}{l c c}
Object & Mass & Radius  \\
       & $(M_\odot)$  & $(R_\odot)$        \\ \hline
& \multicolumn{2}{c}{Masses from binary motion}  \\
Sirius B  &
    $1.000\pm0.016$ & $0.0084\pm0.0002$       \\
Procyon B &
    $0.604\pm0.018$  & $0.0096\pm0.0004$        \\
40 Eri B &
    $0.501\pm0.011$  & $0.0136\pm0.0002$        \\
\hline
\multicolumn{3}{c}{Masses from surface acceleration} \\
EG 50 &
    $0.50\pm0.06$  & $0.0104\pm0.0006$        \\
GD 140 &
    $0.79\pm0.09$  & $0.0085\pm0.0005$        \\
\hline
\multicolumn{3}{c}{Masses from surface gravitational redshift} \\
CD-38 10980 &
    $0.74\pm0.04$  & $0.01245\pm0.0004$        \\
W485A &
    $0.59\pm0.04$  & $0.0150\pm0.001$        \\
\end{tabular}
\end{center}
\end{table}


\newpage
\begin{figure*}
\vspace{-1cm}
\begin{center}
\caption{The mass-radius relation for white dwarfs. 
Solid lines labeled He, C, Mg, and Fe denote the zero-temperature 
mass-radius relation for a star composed of each element of 
Ref.~\cite{HamSal61}. Models of white dwarfs with a hydrogen atmosphere 
\cite{Wood95} are for an effective atmosphere temperature of 
30,000\Kelv (dotted line) and 15,000\Kelv and 
8000 \Kelv (dashed lines). Estimates for observed white dwarfs are shown 
with a $1\sigma$ error. \label{fig:obsWDs} 
[From Provencal et al.~\cite{Provetal98}, 
$\copyright$ 1998. The American Astronomical Society] }
\includegraphics[width=10cm]{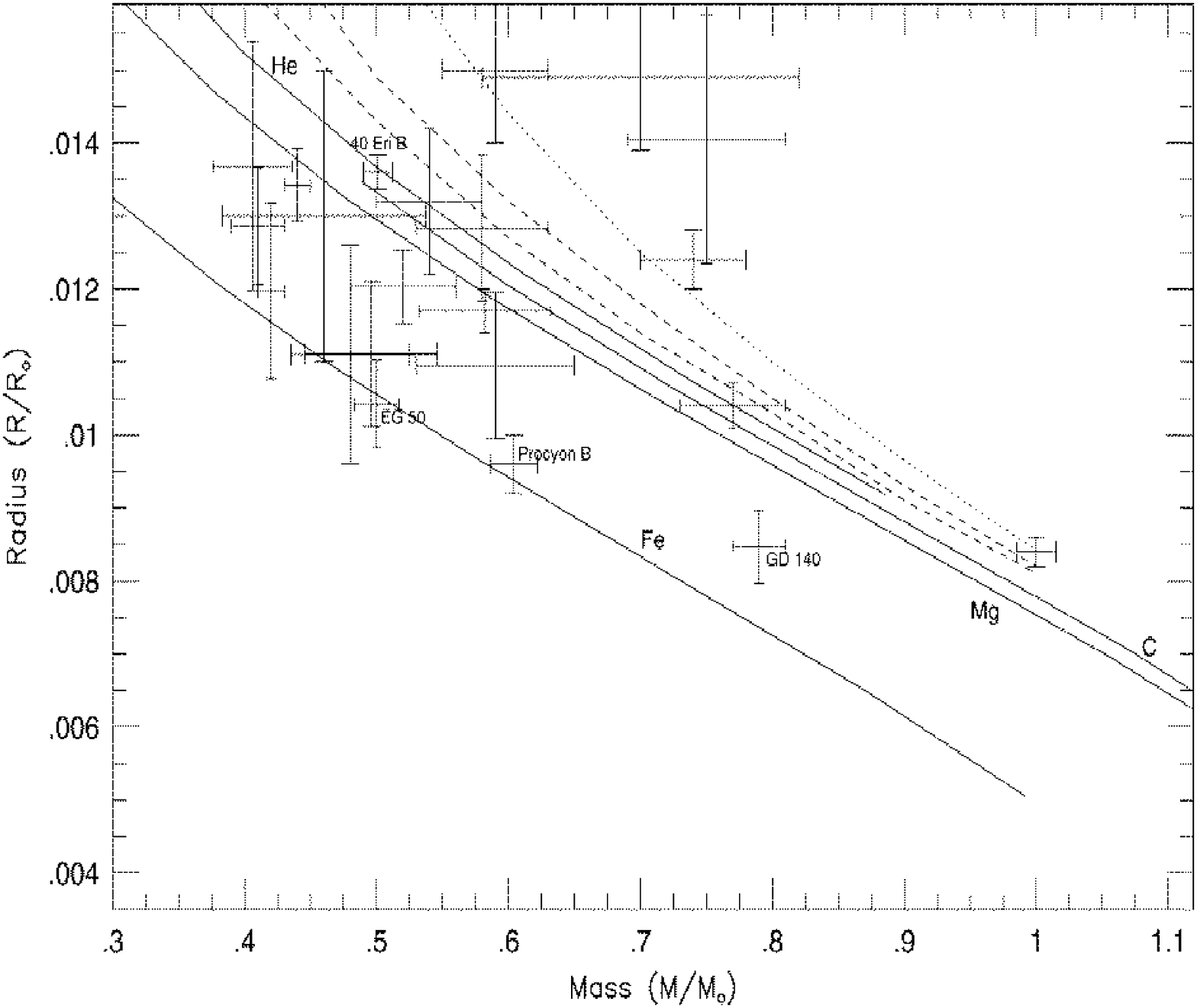}
\end{center}
\end{figure*}

\section{Observations of Condensed Matter in Neutron Stars}
\label{sect:NSs}

Neutron stars are often identified observationally as very accurately  
pulsating sources, where the pulsation is attributed to the star's rotation. 
The period of pulsation limits the size of the source to no more than a few 
tens of kilometers, thus indicating the presence of a very compact 
object. Since the first discovery in 1967, close to 800 pulsars have 
been identified, over 700 of those in radio waves \cite{pulsarcatalog}. 
About thirty X-ray pulsars have also been observed, and a smaller number 
of nonpulsating X-ray sources of various types are also most likely 
to contain a neutron star \cite{XRBbook}.

Theory suggests that neutron stars are formed when a massive star is 
disrupted in a ``supernova'' explosion. As the massive star evolves, matter 
in its core undergoes thermonuclear burning all the way to iron, which is 
the most tightly bound nucleus and therefore cannot burn further. When the 
iron core is massive 
enough it collapses under its own weight, until the collapse is halted when 
nuclear densities are reached. Most of the gravitational binding energy is 
released in the form of neutrinos, but some is transformed into 
an outgoing shock wave that expels the star's envelope and leaves behind 
the newly formed neutron star. Indeed, several dozen young neutron stars 
have been found in sites of past supernovae, confirming this scenario.

\subsection{Neutron Star Masses}

The masses of over twenty neutron stars have been determined observationally 
through their gravitational pull on a binary companion star. 
Within the errors of measurement, the masses of these stars appear to 
cluster in a narrow range 
of $1.35\pm 0.1\;M_\odot$ \cite{ThorChak99}. This is especially true of 
the masses of neutron stars in the four known binary-pulsar systems (where 
the companion object is also a neutron star), listed in Table 
\ref{tab:binmass}, where mass determinations are most accurate. 
Another example is the neutron star - white dwarf binary (once believed 
to be a double binary neutron star system) B2303+46. The properties of the 
orbit allow to limit the mass of the neutron star to 
$1.2<M_{NS}<1.4\;M_\odot$ \cite{vKwKulk99}. 
In particular, the Hulse-Taylor binary pulsar 1913+16 
(the coordinates of its location in the sky) has 
been very accurately determined to have $M_{1913+16}=1.4411\pm0007\;M_\odot$.
Clearly, a high density EoS must satisfy $M_{max}\geq1.44\;M_\odot$ to be 
consistent with this observation.   

\begin{table}[h]
\caption{Mass determinations with errors of neutron stars in known 
binary pulsar systems. [Adapted from Thorsett and Chakrabarty 
\cite{ThorChak99}, $\copyright$ 1999. The American Astronomical Society].
\label{tab:binmass}}
\begin{center}
\begin{tabular}{l c c c}
              &    Median mass   &        68\%    &        95\%    \\
Star   & mass ($M_\odot$) & central limits & central limits \\  \hline 
J1518+4904 pulsar & 1.56 & $+0.13/-0.44$   \\
J1518+4904 companion & 1.05&$+0.45/-0.11$  \\ \hline
B1534+12 pulsar & 1.339 & $\pm 0.003$ & $\pm 0.006$ \\
B1534+12 companion & 1.339 & $\pm 0.003$ & $\pm 0.006$ \\ \hline
B1913+16 pulsar & 1.4411 & $\pm 0.00035$ & $\pm 0.0007$ \\ 
B1913+16 companion &1.3874 & $\pm 0.00035$ & $\pm 0.0007$ \\ \hline 
B2127+11C pulsar& 1.349 & $\pm 0.040$ & $\pm 0.080$  \\ 
B2127+11C companion & 1.363 & $\pm 0.040$ & $\pm 0.080$ \\  \hline
\end{tabular}
\end{center}
\end{table}

This limit does not impose a serious constraint on realistic models for 
the supernuclear EoS and is satisfied by all but the softest equations. 
There has been much debate whether the narrow range of observed neutron 
star masses is evidence for the maximum mass being rather low 
($M_{max}(NS)\approx 1.5\;M_\odot$) \cite{BetheBrown95}, or whether it is 
due to astrophysical effects, which could restrict the range of masses of 
observable stars.  Recently, there is 
growing evidence that heavier neutron stars exist: one in the Vela X-1 
binary is now estimated to have a mass of $\sim1.9\;M_\odot$ \cite{VKerk}, 
and the oscillations observed in some galactic X-ray sources seem to 
indicate that they include neutron stars with masses larger than 
$2\;M_\odot$ \cite{vdKlis}. 
If these estimates are confirmed with improved observations in the future, 
they will provide a much more stringent test that would be consistent only 
with the stiffer models for the supernuclear EoS. 

\subsection{Radii}

The mass and radius relation of a nonrotating 
neutron star is uniquely defined for any given EoS through the solution of 
the TOV equations (Figure \ref{fig:MvsR}). However, to date, 
observational methods for estimating neutron star radii lack the accuracy 
required to critically differentiate between realistic equations of state. 
Hopefully, data from a new generation of X-ray satellites will provide 
substantial constraints on the compactness ($M/R$) of neutron stars (either 
through their thermal emission \cite{Ogelman} or from emission of material 
accreting upon them \cite{MilLamb98}).

\subsection{Rotation periods}

The observed pulsation period of a neutron star is attributed to rotation. 
Most measured periods are in the range $0.25-1\;$sec, but a subclass of 
{\it millisecond} pulsars are also known to exist. The fastest spinning 
neutron star observed to date has a period of $P=1.56\;$msec (so that it 
spins faster than an egg-beater!). 
  
Rotation provides an additional centrifugal barrier which assists the 
internal pressure in supporting the gravitational load. 
The resulting configuration is then dependent on both central density and 
rotation period, and must be solved self-consistently including the effects 
of general relativity. The maximum mass a given equation of state can 
support increases with respect to the static (TOV) value as the star is 
allowed to rotate faster \cite{CST94}. There must exist a lower limit on the 
rotation {\it period} (upper limit on the rotation {\it rate}), since too 
rapid rotation would cause the star to 
shed mass at the equator. The general trend is that a softer EoS predicts a 
smaller lower limit on the rotation period, since a more compact star is more 
tightly bound gravitationally and 
so is better suited to resist the centrifugal force. At present, the limit 
of $\sim1.56\;$msec is not very restrictive, and is consistent with basically 
all realistic models of high density matter. Discovery of a sub-millisecond 
($P\gtrsim 0.5\;$msec) pulsar would allow to distinguish more directly 
between competing supernuclear equations of state \cite{ShaTeu83, STW83, 
Latal90}. 

\subsection{Pulsar Glitches}

Pulsar periods are observed to increase gradually with time, implying that 
rotational energy of the star is being lost (primarily by electromagnetic 
dipole radiation). However, several pulsars have been observed to undergo 
sudden {\it decreases} in the rotation period, which are believed to 
originate from the transfer of angular momentum between different parts of 
the star. The current leading model \cite{PinAlp85,Alparal93} 
suggests that the source of excess angular momentum is the inner curst, and 
thus the magnitude of glitch phenomena can be used to set a lower limit 
on the fraction of the total moment of inertia the inner crust must carry. 
Recent results \cite{LEL99} suggest the crust of the Vela pulsar must 
carry at least $1.4\%$ of the total moment of inertia of the star, which is 
sufficient to rule out equations of state which are very soft in the range 
$1\leq \rho/\rho_{nuc}\leq 2$. Possible progress in theoretical modeling of 
glitch phenomena could provide further limits on the properties of high 
density matter near the nuclear saturation density.

\section{Thermal Properties of Matter in White Dwarfs and Neutron Stars}
\label{sect:Therm}

Fermions in both white dwarfs and neutron stars are expected to have 
degenerate energies which are much larger than their thermal energies. 
However, compact objects do have finite temperatures that are a relic of 
their progenitors. Since degenerate matter has very long scattering 
lengths for individual particles, the cores of both white dwarfs and neutron 
stars are expected to be excellent thermal conductors, and 
practically isothermal. 

Finite temperature effects on the properties of condensed matter in compact 
objects has become a very active theoretical and observational 
field in recent years, especially in the case of white dwarfs. A finite 
temperature leads to cooling through thermal emission processes which are 
potentially observable. Examining the cooling history (i.e., temperature 
vs.~age) of compact objects thus provides empirical evidence 
regarding the thermal properties of high density matter. 
Here we can only briefly mention some of the topics of current interest.

\subsection{Thermal Effects in White Dwarf Structure}

The region most affected by a finite temperature is the low density 
outer-layers of a white dwarf, which may include a thin atmosphere composed 
of helium and possibly also hydrogen. For lower densities the 
electrons are not degenerate and a surface temperature of $10^4\;$\Kelv  
will severely alter the equation of 
state, especially in the regime where electrostatic corrections are 
important. Modern equations of state for a hydrogen/helium atmosphere 
\cite{Saumon95} have been incorporated in several studies of white 
dwarf structure (see \cite{HanPhin98} for helium stars, \cite{Wood95} for 
carbon-oxygen, and \cite{Argen99} for a recent compilation of various 
compositions). The thermal pressure in the outer layers is 
generally found to be very effective in inflating the stellar radius. 
As seen from Figure \ref{fig:obsWDs}, a surface temperature of 
$T_{eff}=10^4\;$\Kelv increases the radius by a significant fraction 
\cite{Wood95, Argen99} with respect to the cold \cite{HamSal61} models.

There has been a recent revival of studies of white 
dwarf thermal evolution, and its dependence on the properties of matter at 
the relevant high densities. We briefly review here the two main aspects of 
these studies, namely cooling and pulsations.

\subsubsection{White Dwarf Cooling}
    
The basic theory of white dwarf cooling was established by Mestel 
in 1952 \cite{Mestel52}. The heat is originally stored in the nondegenerate 
ion-lattice, and the dominant cooling mechanism is photon diffusion to the 
surface from the isothermal core through the nondegenerate outer layers. 
The simplest results are based on Kramer's approximation for the 
photon opacities in the nondegenerate regime:
\begin{equation}\label{eq:Kramer}
\kappa=\kappa_0 \rho T^{-3.5}\;,
\end{equation} 
where $\kappa$ is the mean opacity (in \cmsgm) and 
$\kappa_0=4.34\times 10^{24} Z (1+X_H)\;$\cmsgm is a 
composition dependent constant, $X_H$ and $Z$ being the mass-fraction of 
hydrogen and heavy (not hydrogen or helium) elements, respectively. 
The luminosity through the outer layers, $L(r)$, is dependent on the 
temperature gradient according to the radiative diffusion equation 
\begin{equation}\label{eq:LgradT}
L(r)=-4\pi r^2 \frac{c}{3\kappa \rho}\frac{d}{dr}(aT^4)\;
\end{equation} 
(where $a=7.7565\times 10^{-15}\;\mbox{erg}\;\mbox{deg}^{-4}\;$ is the 
radiation constant). Combining the above thermal temperature gradient with 
the pressure gradient required for hydrostatic stability 
(Eq.~(\ref{eq:TOV}), see \cite{ShaTeu83}), one finds 
that the surface luminosity satisfies
\begin{equation}\label{eq:LofT}
L=5.7\times 10^6\mu Y_e^2 \frac{1}{Z(1+X)}
\left(\frac{M}{M_\odot}\right) T_c^{3.5}\;, 
\end{equation}
where $\mu$ is the mean molecular weight in the nondegenerate atmosphere 
and $T_c$ is the core temperature in the star. For typical white dwarfs 
($M\sim 1\;M_\odot$) with no hydrogen in the atmosphere, 
$X=0,\;Z\approx 0.1$, $\mu=1.4,\; Y_e=0.5$, an observed luminosity in the 
range $10^{-5}-10^{-2}\;L_\odot$ corresponds to a central temperature of 
$T_c\sim10^6-10^7\;$\Kelv.

The key elements of Eq.~(\ref{eq:LofT}) are that the surface luminosity 
(which is observable) is clearly related to the mass of the star, the 
composition and the internal temperature. If the mass is also determined 
separately (as discussed above), luminosity can then provide a direct test 
of the star's thermal and composition profiles. Furthermore, the age of the 
white dwarf can also be estimated based on its cooling time scale, which 
is dependent on the luminosity function (Eq.~\ref{eq:LofT}), the stellar 
mass and the specific heat capacity of a white dwarf material. The latter 
(in \ergsdeggm) if roughly that of the ion lattice, 
\begin{equation}\label{eq:c_v}
c_v(\mbox{ions})=2\frac{3}{2}\frac{k_B}{\mu}\;,
\end{equation}
where the factor of 2 comes from the three degrees of collective vibration 
and rotation, in addition to the three single ion degrees of motion. 
Due to their degeneracy, the specific heat capacity of the electrons is 
suppressed by a factor of $\sim k_B T/E_F$ \cite{Pathria}, 
which for a density of $10^8\;$\gmcmc and an internal temperature of 
$\sim 10^7\;$\Kelv is $10^{-3}$. 
 
While models which include more detailed microphysics do reproduce the 
general results of the Mestel model (see, for example, 
in Refs.~\cite{ShaTeu83, DanMaz90, IbenTut84, Hansen99}), 
accurate work that is compatible with the quality 
of current observations must include several important perturbations. 
Most notably, accurate low temperature opacities, semidegenerate electron 
thermal conductivity and pressure-induced ionization must be accounted for 
in modeling the white dwarf atmosphere. Crystallization of the ion gas 
may also have a significant effect on the thermal history of the star, both 
as a transient source of (latent) heat and by effectively reducing the heat 
capacity. In particular, below the {\it Debye temperature}, 
$\Theta_D\asim 4\times 10^3 \rho^{1/2}\;$ \mbox{\Kelv}, 
the specific heat capacity becomes temperature dependent \cite{ShaTeu83},
\begin{equation}\label{eq:lowTcv} 
c_v=\frac{12 \pi^4}{5} k_B \left(\frac{T}{\Theta_D}\right)^3\;.
\end{equation}  
The specific effects of the finite temperature on the thermodynamic 
properties of matter in white dwarf atmosphere are most notable through 
studies of pulsations (see below). 

It is noteworthy that for several years there seemed to be a paucity of 
low-luminosity (and therefore, old) observed white dwarfs, which posed 
a nontrivial puzzle in terms of models for the stellar evolution. Only 
rather recently it was realized \cite{Hansennat} that the properties of the 
finite temperature atmospheres would make old white dwarfs dimmer than 
previously anticipated (especially if their atmosphere included only a 
helium layer and no hydrogen).  

\subsubsection{Pulsation of White Dwarfs}

The basic theory of stellar pulsation is reviewed in detail by Sarbani Basu 
in this volume. We note that studies of pulsations in general allow one to 
estimate the sound speed profile of the star, 
its temperature gradient, and even the mean molecular weight in its core. 
It has been well established \cite{Coxbook} that stellar 
pulsation provides an effective probe for the properties of matter 
in a star. Commonly referred to as ``asteroseismology'', this field is 
rapidly emerging due significant advances in observational instrumentation 
and theoretical modeling. 

White dwarfs, like many other stars, may undergo stable pulsations which 
leave a detectable imprint on the time dependence of the star's luminosity 
\cite{WDastseiss, BradWin91}. 
Since pulsations tend to damp over time (or the star evolves out of the 
instability strip, if it is overstable to the pulsations), they are most 
easily observable in younger, and therefore hotter, white dwarfs. 
Modeling observed pulsations of young white dwarfs is particularly 
important for probing the EoS of semidegenerate matter 
\cite{BradWin92,MonWin99}. 
With the aid of theoretical models, studies of 
white dwarfs pulsations have been used for independently restricting the 
mass-radius relationships and dimensions of the hydrogen 
envelope \cite{BradWin91, GoldWu99}, and in general seems to hold much 
potential for future studies of the details of white dwarf physics. For 
example, some pulsation modes are theoretically predicted to be more 
sensitive to the extent of crystallization of the atmosphere material. 
Recent analysis of the periods and relative magnitudes various 
pulsation modes in the white dwarf BPM 37093 \cite{MonWin99} helps determine  
the extent of crystallization in the atmosphere (about 50\% in mass), which 
compares favorably with theory. In another recent example \cite{Costa99}, 
it is possible to identify, for the first time, the contraction rate of the 
pre-white dwarf star PG 1159-035, by measuring the secular changes over time 
of the periods of various modes.    

\subsection{Thermal Effects in Neutron Star Structure}

Although finite temperature effects are practically negligible when 
considering the overall structure of evolved neutron stars, they are 
important in assessing their thermal history and as probes of matter in 
the interior of these objects. Some thermal effects on structure do exist 
in newly-born neutron stars (often called ``proto-neutron star'').

\subsubsection{Proto-Neutron Stars}

Neutron stars are born in supernovae with initial internal temperatures of 
several times $10^{11}\;$\Kelv and initial entropies of $\sim 2 k_B$ per 
nucleon ($k_B$ is the Boltzmann constant). These values are large enough to 
impose nontrivial effects on the composition and structure of the neutron 
star compared to the $T=0$ case \cite{Prakal97}. At such high 
temperatures the matter is {\it not} transparent to the thermal neutrinos 
\cite{ShaTeu83, RedPrak97}, and the proto-neutron star must cool through 
neutrino diffusion to the surface, which occurs over a time scale of several 
seconds (much longer than a dynamical time of milliseconds) \cite{BurLat86}. 
This fundamental prediction was dramatically confirmed by the observed 
neutrino pulse from the supernova 1987A (the closest supernova 
observed from the Earth in 400 years), which lasted for $\sim 12$ seconds. 
(see \cite{Bethe90} for a review). Further studies of the properties of high 
temperature, supernuclear-density matter are required to model this brief 
early cooling epoch and its possible observational signatures \cite{Ponsal99}.
    
\subsubsection{Cooling of Neutron Stars}

After the initial rapid cooling stage which lasts several days, a neutron 
star settles on a slower cooling curve \cite{ShaTeu83, Pethick92}. 
Once neutrinos escape the system freely, they must be continuously produced 
in order to fuel the emission process. Subsequent neutrinos are mostly 
produced by the so-called URCA ($\beta$ and inverse $\beta$ decays as in 
Eq.~(\ref{eq:betas})) and other processes. The isothermal core at this stage 
is expected to have temperature of several times $10^8\;$\Kelv, while at the 
surface the temperature is down to several times $10^6\;$\Kelv - which 
provides for continuous surface thermal luminosity in soft X-rays.
This soft X-ray luminosity is much more difficult to detect 
than the optical ultraviolet luminosity white dwarfs. Current observations 
are generally limited to young (age $\lesssim10^6\;$yrs) neutron stars, where 
the main cooling mechanism is still the neutrino emission from the core. 
An important consequence of this situation is that estimates of thermal 
emission from neutron stars serve as a probe of the properties of matter in 
the core \cite{Schaabal96}.

Over 20 compact, isolated soft X-ray sources observed with X-ray 
imaging telescopes have been identified as neutron stars \cite{SchaabWW}. 
In most cases the source 
is either too faint to allow for a spectroscopic determination of the surface 
temperature, or the emission is dominated by other phenomena (most likely 
magnetospheric emission). However, in a hand-full of cases surface thermal 
emission has been strong enough to be detected and the surface temperature 
has been deduced to within a factor of two \cite{SchaabWW}. 
The measurements are still not 
accurate enough to determine whether the observed layer is the actual surface 
or a possible hydrogen atmosphere, which would have different emission 
properties, but they are sufficient to place some constraints on the 
rates of neutrino emitting processes in the core \cite{Page98}. 

Perhaps the most notable conclusion to date of observed neutron star cooling 
rates is that they support the theoretical prediction that the nucleons in 
the core couple to a {\it superfluid} state. Theoretical models suggest that 
the strong interactions will pair the neutrons in the core in a $^3P_1$ 
superfluid and the protons to a $^1S_0$ superconductor with critical 
temperatures of the order of $10^9\;$\Kelv \cite{Takatsuka}. Neutrino 
emitting processes, such as $\beta$-decays must break a coupled Cooper pair 
before its constituents can participate in the decay. The main effect of 
nucleon superfluidity is therefore damping the efficiency of neutrino 
emission from the core (there is also a secondary effect due to the 
modulation of the heat capacity, which first increases discontinuously as the 
star cools to the critical temperature, and then decreases exponentially at 
lower temperatures). Observed neutron star surface temperatures are 
apparently too high to be consistent with the cooling rates predicted for 
normal core, but they agree with the suppressed cooling rates when the 
nucleons couple to a superfluid state \cite{SchaabSF,SchBalSch98}. 
Further progress in analyses of the matter in the interiors of neutron stars 
is expected through measurements by NASA's recently launched Chandra X-ray 
satellite. 

\section{Concluding Remarks}\label{sect:CONC}

The theoretical and observational study of compact objects remains one of 
the most exciting fields in modern astronomy. In essence, this research is 
also an exploration of the properties of condensed matter at extreme 
densities. Predictions regarding the properties of white dwarfs and neutron 
stars serve to test our understanding of matter at these high densities, 
while theories of high density matter serve as a basis for interpreting 
observational results regarding these objects. Most exciting, these objects 
bring together all four of the fundamental forces of nature and probe 
regimes not accessible in the terrestrial laboratory. They provide the most 
numerous and accessible sample of objects where relativistic gravitation - 
general relativity - plays a role in determining their physical properties.

In this review we have described the tight interconnection 
between the microscopic (local) properties of condensed matter at high 
densities and the macroscopic (global) properties of white dwarfs and 
neutron stars. 
While the fundamental principles of cold, high density matter are 
believed to be well understood, and are generally consistent with 
observations, key questions still remain, and new observations may give rise 
to new puzzles. The current boom in capabilities of Earth-bound 
telescopes and satellite instrumentation promises that many more puzzles - 
and hopefully, answers - are 
in store regarding the nature of cosmic matter at high densities.

\def\vol#1{{\bf #1}}
\def\aap{{\it Astron.~Astrophys.}}
\def\aj{{\it Astron.~J.}}
\def\apj{{\it Astrophys.~J.}}
\def\apjl{{\it Astrophys.~J.~Lett.}}
\def\apjs{{\it Astrophys.~J.~Supp.}}
\def\araa{{\it Ann.~Rev~Astron.~Astrophys.}}
\def\arnp{{\it Ann.~Rev~Nuc.~Pat.~Sci.}}
\def\nat{{\it Nature}}
\def\mnras{{\it Mon.~Not.~Roy.~Astron.~Soc.}}
\def\nphysa{{\it Nucl.~Phys.~A}}
\def\physletb{{\it Phys.~Lett.~B}}
\def\physrep{{\it Phys.~Rep.}}
\def\physrev{{\it Phys.~Rev.}}
\def\prc{{\it Phys.~Rev.~{\bf C}}}
\def\prl{{\it Phys.~Rev.~Lett.}}
\def\rmphys{{\it Rev.~Mod.~Phys}}


\begin{thebibliography}{99}

\bibitem{ShaTeu83} Shapiro, S.~L., and Teukolsky, S.~A.~(1983).
        {\it ``Black Holes, White Dwarfs and Neutron Stars''}. 
        Wiley, New York, NY.

\bibitem{BPS}
        Baym, G., Pethick, C.~J., and Sutherland, P.~(1971). 
        The ground state of matter at high densities: Equation of state and 
stellar models. \apj, \vol{170}, 299.

\bibitem{ChandraWD} 
        Chandrasekar, S.~(1931).
        The density of white dwarf stars. {\it Phil.~Mag.} {\bf 11}, 592; 
        The maximum mass of ideal white dwarfs. \apj \vol{74}, 81. 

\bibitem{Sal61} Salpeter, E.~E.~(1961).
        Energy and pressure of zero temperature plasma. 
        \apj, \vol{134}, 669.

\bibitem{Dirac} Dirac, P.~.A.~M.~(1926).
        On the theory of quantum mechanics. 
        {\it Proc.~Roy.~Soc.~London Ser.~A}, \vol{112}, 661.

\bibitem{FMT}
        Feynman, R.~P., Metropolis, N., and Teller, E.~(1949).
        Equations of state of elements based on the generalized Fermi-Thomas 
Theory. \physrev, \vol{75}, 1561.

\bibitem{HW} Harrison, B.~K., and Wheeler, J.~A.~(1958).
        See {\it in} Harrison, B.~K., Thorne, K.~S., and 
        Wheeler, J.~A.~(1965). Gravitation theory and gravitational collapse. 
        University of Chicago Press, Chicago.

\bibitem{BBP}
        Baym, G., Bethe, H.~A., and Pethick, C.~J.~(1971). 
        Neutron star matter. \nphysa, \vol{175}, 225.

\bibitem{RavPeth} 
        Ravenhall, D.~G., and Pethick, C.~J.~(1995).
        Matter at large neutron excess and the physics of neutron-star crusts.
        \arnp, \vol{45}, 429.

\bibitem{WFF} Wiringa, R.~B., Fiks, V., and Fabrocini, A.~(1988). 
        Equation of state for dense nuclear matter.
        \prc, \vol{38}, 1010

\bibitem{APR} Akmal, A., Pandharipande, V.~R. and Ravenhall, D.~G. (1998).    
        The equation of state for nucleon matter and neutron star structure.
        \prc\vol{58}, 1804.

\bibitem{SerWal79}
        Serot, B.~ D., and Walecka, J.~D.~(1979). 
        Properties of finite nuclei in a relativistic quantum field theory. 
        \physletb, \vol{87}, 172.

\bibitem{Glenbook} 
        Glendenning, N.~K~(1996).
        ``{\it Compact Stars}'', Springer, New York, NY.

\bibitem{LatSwes}
        Lattimer, J.~M., and Swesty, D.~F.~(1991). 
        An effective equation of state for hot dense matter.
        \nphysa, 535, 331  

\bibitem{BG}
        Balberg, S., and Gal, A.~(1997). 
        An effective equation of state for dense matter with strangeness.
        \nphysa, \vol{625}, 435.

\bibitem{Prakal97}
        Prakash, M.~et~al.~(1997). 
        Composition and structure of proto-neutron stars.
        \physrep, \vol{280}, 1.

\bibitem{BLC}
        Balberg, S, Lichtenstadt, I., and Cook, G.~B.~(1999).
        Roles of hyperons in neutron stars. \apjs, \vol{121}, 515.

\bibitem{HamSal61}
       Hamada, T., and Salpeter, E.~E.~(1961).
       Models for zero-temperature stars, \apj, \vol{134}, 683.

\bibitem{Tolman}
        Tolman, R.~C.~(1939).
        Static solutions of Einstein's filed equations for spheres of fluids. 
        \physrev, \vol{55}, 364.

\bibitem{OV}
        Oppenheimer, J.~R., and Volkoff, G.~M.~(1939). 
        On massive neutron cores. \physrev, \vol{55}, 374.

\bibitem{Landau32} 
        Landau, L.~D.~(1932).
        On the theory of stars. {\it Phys.~Z.~Sowjwtunion}, \vol{1}, 285. 


\bibitem{RHF8}
        Huber, H., Weber, F., Weigel, M.~ K., and Schaab, Ch.~(1998).
         Neutron star properties with relativistic equations of state.
        {\it Int.~J.~Mod.~Phys.~E}, \vol{7}, 301.

\bibitem{MF}
        Pandharipande, V.~R., and Smith, R.~A.~(1975).
        Nuclear matter calculations with mean stellar fields.
        {\it Phys.~Lett.~B}, \vol{59}, 15.

\bibitem{FPS}
        Friedman B., and Pandharipande, V.~R.~(1981).
        Hot and cold, nuclear and neutron matter. \nphysa, \vol{361}, 502.

\bibitem{PPS76}
        Pandharipande, V.~R., Pines, D., and Smith, R.~A.~(1976). 
        Neutron star structure: theory, observation and speculation.
        \apj, \vol{208}, 550.

\bibitem{WDcataloge} 
        McCook, G.~P., Sion, E.~M.~(1999).
        A catalog of spectroscopically identified white dwarfs. 
        \apjs, \vol{121}, 1.

\bibitem{Provetal98}
        Provencal, J.~L., Shpiman, H.~L., H{\o}g, E., and Thejll, P.~(1998).
        Testing the white dwarf mass-radius relation with {\it HIPPARCOS}. 
        \apj, \vol{494}, 759.

\bibitem{Ship79}
        Shipman, H.~L.~(1979).
        Masses and Radii of white dwarfs III. Results for 110 hydrogen-rich
and 28 Helium-rich stars.
        \apj, \vol{228}, 240. 

\bibitem{Schmidt96}
        Schmidt, H.~(1996).
        The empirical white dwarf mass-radius relation and its possible 
        improvement by HIPPARCOS.
        \aap, \vol{311}, 852.

\bibitem{ShaTeu76}
        Shapiro, S.~L., and Teukolsky, S.~A.~(1976).
        On the maximum gravitational redshift of white dwarfs.
        \apj, \vol{203}, 697.

\bibitem{Weidermann90}
        Weidermann, V.~(1990)
        Masses and evolutionary status of white dwarfs and their progenitors. 
        \araa, \vol{28}, 103.

\bibitem{DanMaz90}
        D'Antona, F., and Mazzitelli, I.~(1990).
        Cooling of white Dwarfs.        
        \araa, \vol{28}, 139.

\bibitem{Greenstein86}
        Greenstein J,~L.~(1986).
        White dwarfs in wide binaries. I - Physical properties. 
        II - Double degenerates and composites.
        \aj, 92, 859.     

\bibitem{SiriusB}
        Gatewood, G.~D., and Gatewood, C.~V.~(1978).
        A study of Sirius.
        \apj, \vol{225}, 191.

\bibitem{Wood95}
        Wood, M.~A.~(1995). 
        Theoretical white dwarf luminosity functions: DA models. 
        {\it In} ``Proc.~9th European Workshop on White Dwarfs'' 
        (D.~Koster and K.~Werner, Eds.), pp.~41. Springer, Berlin.

\bibitem{pulsarcatalog} 
        Taylor, J.~H., Manchester, R.~N., and Lyne, A.~G.~(1993).
        Catalog of 558 pulsars. 
        \apjs, \vol{88}, 529.
        See the Princeton Pulsar-Group website for an updated catalog, 
        {\tt http://www.pulsar.princeton.edu}.
   
\bibitem{XRBbook}
        {\it X-Ray Binaries}, W.~H.~G.~Lewin, J.~van Paraijs and 
        E.~P.~J.~van den Heuvel, Eds.~(1995). Cambridge University Press, 
        Cambridge.
 
\bibitem{ThorChak99} 
        Thorsett, S.~E., and Chakrabarty, D.~(1999).
        Neutron star mass measurements. I. Radio pulsars.
        \apj, \vol{512}, 288.

\bibitem{vKwKulk99}
        van Kerkwijk, M.~H., and Kulkarni S.~R.~(1999).
        A massive white dwarf companion to the eccentric binary pulsar 
        system PSR B2303+46.
        \apjl, \vol{516}, L25.

\bibitem{BetheBrown95} 
        Bethe, H.~E., and Brown, G.~E.~(1995)
        Observational constraints on the maximum neutron star mass.
        \apjl, \vol{445}, L29.

\bibitem{VKerk} van Kerkwijk, M.~H.~(2000)
       Neutron Star Mass Determinations. 
       {\it To appear in} ``Proc.~ESO Workshop on Black Holes in Binaries and 
Galactic Nuclei", (L.~Kaper, E.~P.~J.~van den Heuvel, and P.~A.~Woudt, Eds.). 
       Springer-Verlag, Berlin (available on the Los-Alamos pre-print server, 
       {\tt astro-ph/0001077}).

\bibitem{vdKlis}
        van der Klis, M.~(1998)
        KlioHertz quasi-periodic oscillations in low mass X-ray binaries. 
        {\it In} ``The Many Faces of Neutron Stars'', NATO ASI series
        (R.~Buccheri, J.~van Paradijs and M.~A.~Alpar, Eds.), pp.~337.
        Kluwer, Dordrecht. 

\bibitem{Ogelman}
        \"Ogelman, H.~(1995). 
        X-ray observations of cooling neutron stars.
        {\it In} ``The Lives of the Neutron Stars'' 
        (M.~A.~Alpar, \"U.~Kizilogu and J.~van Paradijs, Eds.), pp.~101. 
        Dordechet, Kluwer.

\bibitem{MilLamb98}
       Miller, M.~C., and Lamb, F.~K.~(1998).
       Bounds on the compactness of neutron stars from brightness 
       oscillations during X-Ray bursts. \apjl, \vol{499}, L37.

\bibitem{CST94}
        Cook, G.~B., Shapiro, S.~L., and Teukolsky, S.~A.~(1994). 
        Rapidly rotating neutron stars in general relativity: Realistic 
        equations of state
        \apj, \vol{424}, 823.

\bibitem{STW83}
        Shapiro, S.~L., Teukolsky, S.~A., and Wasserman I.~(1983).
        Implications of the millisecond pulsar for neutron star models. 
        \apj, \vol{272}, 702.      

\bibitem{Latal90} 
        Lattimer, J.~M., Prakash, M., Masak, D., and Yahil, A.~(1990).
        Rapidly rotating pulsars and the equation of state.
        \apj, \vol{355}, 241

\bibitem{PinAlp85} 
        Pines, D., and Alpar, M.~A.~(1985). 
        Superfluidity in neutron stars. \nat, 316, 27.

\bibitem{Alparal93}
        Alpar, M.~A., Chau, H.~F., Cheng, K.~S.,and Pines, D.~(1993). 
        Postglitch relaxation of the VELA pulsar after its first eight large 
        glitches - A reevaluation with the vortex creep model.
        \apj, 409, 345.

\bibitem{LEL99}
        Link, B., Epstein, R.~I., and Lattimer, J.~M.~(1999).
        Pulsar constraints on neutron star structure and equation of state.
        \prl, \vol{83}, 3362.

\bibitem{Saumon95}
        Saumon, D., Chabrier, G., and van Horn, H.~M.~(1995).
        An equation of state for low-mass stars and giant planets.
        \apjs, \vol{99}, 713.

\bibitem{HanPhin98}
        Hansen, B.~M.~S., and Sterl Phinney, E.~(1998).
        Stellar forensics. I - Cooling curves. \mnras, \vol{294}, 557.
          
\bibitem{Argen99}
        Panei, J.~L., Althaus, L.~G., and Benvenuto, O.~G.~(2000).
        Mass-radius relations for white dwarfs of different internal 
        compositions. \aap, \vol{353}, 970.

\bibitem{Mestel52}
        Mestel, L.~(1952).
        On the theory of white dwarfs stars I. The energy sources of 
        white dwarfs. \mnras, \vol{112}, 583.

\bibitem{Pathria}
        Pathria, R.~K.~(1972).
        {\it Statistical Mechanics}. Pergamon Press, Oxford.

\bibitem{IbenTut84} 
        Iben, I., Jr., and Tutukov, A.~V.~(1984).
        Cooling of low-mass carbon-oxygen dwarfs from the planetary nucleus 
        stage through the crystallization stage. \apj, \vol{282}, 615.

\bibitem{Hansen99}
        Hansen, B.~M.~S.~(1999).
        Cooling models for old white dwarfs. \apj, \vol{520}, 680.

\bibitem{Hansennat}
        Hansen, B.~M.~S.~(1998).
        Old and blue white-dwarf stars as a detectable source of 
        microlensing events. \nat, \vol{394}, 860.

\bibitem{Coxbook}
        Cox, J.~P.~(1980).
        {\it Theory of Stellar Pulsation}. Princeton University Press, 
        Princeton, NJ.

\bibitem{WDastseiss}
        Bradley, P.~A., Winget, D.~E., and Wood, M.~A.~(1993).
        The potential for asteroseismology of DB white dwarf stars.
        \apj, \vol{406}, 661.

\bibitem{BradWin91}
        Bradley, P.~A., and Winget, D.~E.~(1991).
        Asteroseismology of white dwarf stars. I - Adiabatic results.
        \apjs, \vol{75}, 463.

\bibitem{BradWin92}
        Bradley, P.~A., Winget, D.~E., and Wood, M.~A.~(1993)
        Maximum rates of period change for DA white dwarf models with 
        carbon and oxygen cores. \apjl, \vol{391}, 33.

\bibitem{MonWin99}
        Montgomery, M.~H., and Winget, D.~E.~(1999). 
        The effect of crystallization on the pulsations of white dwarf stars.
        \apj, \vol{526}, 976.

\bibitem{GoldWu99}
        Goldreich, P., and Wu, Y.~(1999).
        Gravity modes in ZZ Ceti stars. I. Quasi-adiabatic analysis of 
        over-stability. \apj, \vol{511}, 904;
        Gravity modes in ZZ Ceti stars. II. eigenvalues and eigenfunctions.
        \apj, \vol{519}, 783.

\bibitem{Costa99}
        Costa, J.~E.~S., Kepler, S.~O., and Winget, D.~E.~(1999).
        Direct measurement of a secular pulsation period change in the 
        pulsating hot pre-white dwarf PG 1159-035. \apj, \vol{522}, 973.

\bibitem{RedPrak97}
        Reddy, S., and Prakash, M.~(1997). 
        Neutrino scattering in a newly born neutron star.
        \apj, \vol{423}, 689.

\bibitem{BurLat86} 
        Burrows, A., and Lattimer, J.~M.~(1986).
        The birth of neutron stars,. \apj, \vol{307}, 178.

\bibitem{Bethe90}~
        Bethe, H.~E.~(1990). 
        Supernovae. \rmphys, \vol{62}, 801.

\bibitem{Ponsal99}
        Pons, J.~A., Reddy, S., Prakash, M., Lattimer, J.~M., and 
        Miralles, J.~A.~(1999).
        Evolution of Proto-Neutron Stars. \apj, \vol{513}, 780.

\bibitem{Pethick92} 
        Pethick, C.~J.~(1992).
        Cooling of neutron stars. \rmphys, \vol{64}, 1133.

\bibitem{Schaabal96}
        Schaab, Ch., Weber, F., Weigel M.~K., and Glendenning, N.~K.~(1996).
        Thermal evolution of compact stars.
        \nphysa, 605, 531.

\bibitem{SchaabWW}
        Schaab, Ch., Weber, F., and Weigel, M. K.~(1998).
        Neutron superfluidity in strongly magnetic interiors of neutron 
        stars and its effect on thermal evolution. \aap, \vol{335}, 596.

\bibitem{Page98}
        Page D.~(1998).
        Thermal evolution of isolated neutron stars. 
        {\it In} ``The Many Faces of Neutron Stars'', NATO ASI series
        (R.~Buccheri, J.~van Paradijs and M.~A.~Alpar, Eds.), pp.~539. 
        Kluwer, Dordrecht. 

\bibitem{Takatsuka}
        Takatsuka, T.~ and Tamagaki, R.~(1993).
        Superfluidity in neutron star matter and symmetrical nuclear matter.
        {\it Prog.~Theo.~Phys.~Suppl.}, \vol{112}, 27.

\bibitem{SchaabSF}
        Schaab, Ch., Voskresensky, D., Sedrakian, A.~D.,Weber, F., 
        and Weigel, M.~K.~(1997).
        Impact of medium effects on the cooling of nonsuperfluid and 
        superfluid neutron stars.
         \aap, 321, 591. 

\bibitem{SchBalSch98}
        Schaab, Ch., Balberg, S., and Schaffner-Bielich, J.~(1998).
        Implications of hyperon superfluidty for neutron star cooling.
        \apjl, 504, L99.

\end{thebibliography}
\end{document}